\documentclass[journal]{IEEEtran}

\newif\ifnewtext
\newtexttrue		

\usepackage{ifpdf}
\usepackage{amssymb}
\usepackage{graphicx}
\usepackage{epstopdf}
\usepackage[cmex10]{amsmath}
\usepackage{url}
\usepackage[linesnumbered,ruled,vlined]{algorithm2e}
\SetKwInput{KwIn}{Initialization}
\usepackage{multirow}
\usepackage{subfigure}
\usepackage{xspace}

\newcommand{\comment}[1]{}

\newcommand{\ourproposal}{TOFEC\xspace}
\newcommand{\ourscheme}{TOFEC\xspace}

\newcommand{\lenVec}{\hat{n}}
\newcommand{\dimVec}{\hat{k}}

\newcommand{\compVec}{\hat{p}}
\newcommand{\rateVec}{\hat{r}}

\newcommand{\ones}{\bf{1}}
\newcommand{\CapSta}{C_{sta}}
\newcommand{\CapFull}{C}
\newcommand{\normArrival}{\overline{\lambda}}
\newcommand{\aveUsage}{\overline{U}}
\newcommand{\fixedDelta}{\overline{\Delta}}
\newcommand{\linearDelta}{\widetilde{\Delta}}
\newcommand{\fixedExp}{\overline{\Psi}}
\newcommand{\linearExp}{\widetilde{\Psi}}

\newcommand{\optSolution}{$(\dimVec^*,\rateVec^*)$\xspace}
\newcommand{\FR}{\Gamma}
\newcommand{\KR}{\Omega}
\newcommand{\RR}{\Upsilon}

\newcommand{\nthreshold}{H^{N}}
\newcommand{\kthreshold}{H^{K}}
\newtheorem{theorem}{\textbf{Theorem}}

\newtheorem{corollary}{\textbf{Corollary}}

\newcommand{\onewidth}{0.74\columnwidth}
\newcommand{\twowidth}{0.45\columnwidth}
\newcommand{\threewidth}{0.3\textwidth}

\newcommand{\shrinkbeforecaption}{-5pt}
\newcommand{\shrinkaftercaption}{-14pt}

\newcommand{\mybox}[1]{\vspace{5pt}\centerline{\framebox{\parbox[c]{0.95\columnwidth}{#1}}}\vspace{5pt}}

\begin{document}

\title{On Throughput-Delay Optimal Access to Storage Clouds via Load Adaptive Coding and Chunking}

\author{Guanfeng~Liang,~\IEEEmembership{Member,~IEEE,}
        and~Ula\c{s}~C.~Kozat,~\IEEEmembership{Senior Member,~IEEE}
\thanks{G. Liang and U.C. Kozat are with DOCOMO Innovations Inc., Palo Alto, California USA. G. Liang is the contact author. E-mail: gliang@docomoinnovations.com}
}
\maketitle

\begin{abstract}
Recent literature including our past work provide analysis and solutions for using (i) erasure coding, (ii) parallelism, or (iii) variable slicing/chunking (i.e., dividing an object of a specific size into a variable number of smaller chunks) in speeding up the I/O performance of storage clouds.  However, a comprehensive approach that considers all three dimensions together to achieve the best throughput-delay trade-off curve had been lacking.  This paper presents the first set of solutions that can pick the best combination of coding rate and object chunking/slicing options as the load dynamically changes. Our specific contributions are as follows: (1) We establish via measurements that combining variable coding rate and chunking is mostly feasible over a popular public cloud. (2) We relate the delay optimal values for chunking level and code rate to the queue backlogs via an approximate queuing analysis. (3) Based on this analysis, we propose \ourscheme that adapts the chunking level and coding rate against the queue backlogs. 
\comment{Under light workloads, \ourscheme creates  smaller chunks and uses more parallel connections per file, minimizing service delay. Under heavy workloads, \ourscheme automatically reduces the level of chunking (fewer chunks with increased size) and uses fewer parallel connections to reduce overhead, resulting in higher throughput and preventing queueing delay.} 
Our trace-driven simulation results show that \ourscheme's adaptation mechanism converges to an appropriate code that provides the optimal throughput-delay trade-off without reducing system capacity. Compared to a non-adaptive strategy optimized for throughput, \ourscheme delivers $2.5\times$ lower latency under light workloads; compared to a non-adaptive strategy optimized for latency, \ourscheme can scale to support over $3\times$ as many requests. (4) We propose a simpler greedy solution that \comment{does not require the modeling of the delay distribution. The greedy solution} 
performs on a par with \ourscheme in average delay performance, but exhibits significantly more performance variations.   
\comment{
}

\comment{
\ourscheme is a strategy that helps front-end proxy adapt to different levels of workload by treating scalable cloud storage (e.g. Amazon S3) as a shared resource requiring admission control. Under light workloads, \ourscheme creates more smaller chunks and uses more parallel connections per file, minimizing service delay. Under heavy workloads, \ourscheme automatically reduces the level of chunking (fewer chunks with increased size) and uses fewer parallel connections to reduce overhead, resulting in higher throughput and preventing queueing delay. Our trace-driven simulation results show that \ourscheme's adaptation mechanism converges to an appropriate code that provides the optimal throughput-delay trade-off without reducing system capacity. Compared to a non-adaptive strategy optimized for throughput, \ourscheme delivers $2.5\times$ lower latency under light workloads; compared to a non-adaptive strategy optimized for latency, \ourscheme can scale to support over $3\times$ as many requests.
}
\end{abstract}


\IEEEpeerreviewmaketitle

\section{Introduction}
\label{sec:intro}

Cloud storage has gained wide adoption as an economic, scalable, and reliable mean of providing data storage tier for applications and services. 
Typical cloud storage systems are implemented as key-value stores in which data objects are stored and retrieved via their unique keys. To provide high degree of availability, scalability, and data durability, each object is replicated several times within the internal distributed file system and sometimes also further protected by erasure codes to more efficiently use the storage capacity while attaining very high durability guarantees \cite{Huang12}. 

Cloud storage providers usually implement a variety of optimization mechanisms such as load balancing and caching/prefetching internally to improve performance. Despite all such efforts, still evaluations of large scale systems indicate that there is a high degree of randomness in delay performance \cite{Garfinkel07anevaluation}. 
Thus, services that require more robust and predictable Quality of Service (QoS) must deploy their own external solutions such as sending multiple/redundant requests (in parallel or sequentially), chunking large objects into smaller ones and read/write each chunk through parallel connections, replicate the same object using multiple distinct keys in a coded or uncoded fashion, etc.

In this paper, we present  {\bf black box} solutions\footnote{They use only the API provided by storage clouds and do not require any modification or knowledge of the internal implementation of storage clouds.} that can provide much better throughput-delay performance for reading and writing files on cloud storage utilizing (i) parallelism, (ii) erasure coding, and (iii) chunking. To the best of our knowledge, our work is the first one that adaptively picks the best erasure coding rate and chunk size to minimize the expected latency without sacrificing the supportable rate region (i.e., maximum requests per second) of the storage tier.  The presented solutions can be deployed over a proxy tier external to the cloud storage tier or can be utilized internally by the cloud provider to improve the performance of their storage services for all or a subset of their tenants with higher priority.


\begin{figure}[!t]
\centering
\includegraphics[width = \onewidth]{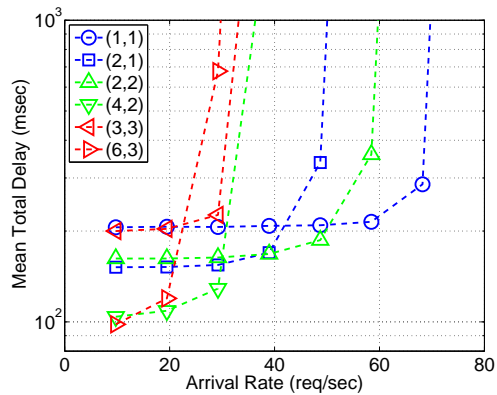}
\caption{Delay for downloading 3MB files using fixed MDS codes}
\label{fig:fixedDelays}
\end{figure}

\subsection{State of the Art}
Among the vast amount of research on improving cloud storage system's delay performance emerged in the past few years, two groups in particular are closely related to our work presented in this paper:

{\bf Erasure Coding with Redundant Requests:} As proposed by authors of \cite{fastcloud,Longbocodeingincloud, MDS-queue}, files (or objects) are divided into a {\em pre-determined} number of $k$ chunks, each of which is $1/k$ the size of the original file, and encoded into $n>k$ of ``coded chunks'' using an $(n,k)$ Maximum Distance Separable (MDS) code, or more generally a Forward Error Correction (FEC) code. Downloading/uploading of the original file is accomplished by downloading/uploading $n$ coded chunks using parallel connections simultaneously and is deemed served when download/upload of any $k$ coded chunks complete. Such mechanisms significantly improves the delay performance under light workload. However, as shown in our previous work \cite{fastcloud} and later reconfirmed by \cite{MDS-queue}, system capacity is reduced due to the overhead for using smaller chunks and redundant requests. This phenomenon is illustrated in Fig.\ref{fig:fixedDelays} where we plot the throughput-delay trade-off for using different MDS codes from our simulations using delays traces collected on Amazon S3. Codes with different $k$ are grouped in different colors. Using a code with high level of chunking and redundancy, in this case a $(6,3)$ code, although delivers $2\times$ gain in delay at light workload, reduces system capacity to only $30\%$ of the original basic strategy without chunking and redundancy, i.e., $(1,1)$ code!

This problem is partially addressed in \cite{fastcloud} where we present strategies that adjust $n$ according to workload level so that it achieves the near-optimal throughput-delay trade-off for a {\em predetermined} $k$. For example, if $k=3$ is used, the strategies in \cite{fastcloud} will achieve the lower-envelope of the red curves in Fig.\ref{fig:fixedDelays}. Yet, it still suffers from an almost 60\% loss in system capacity.

{\bf Dynamic Job Sizing:}
It has been observed in \cite{Garfinkel07anevaluation, stout} that in key-value storage systems such as Amazon S3 and Microsoft's Azure Storage, throughput is dramatically higher when they receive a small number of storage access requests for large jobs (or objects) than if they receive a large number of requests for small jobs (or objects), because each storage request incurs overheads such as networking delay, protocol-processing, lock acquisitions, transaction log commits, etc. Authors of \cite{stout} developed Stout in which requests are dynamically batched to improve throughput-delay trade-off of key-value storage systems. Based on the observed congestion Stout increase or reduce the batching size. Thus, at high congestion, a larger batch size is used to improve the throughput while at low congestion a smaller batch size is adopted to reduce the delay.

\subsection{Main Contributions}
Our work unifies the ideas of redundant requests with erasure coding and dynamic job sizing together in one solution framework. Our major contributions can be listed as follows.
\begin{itemize}
\item Providing dynamic job sizing while maintaining parallelism and erasure coding gains is a non-trivial undertaking. Key-value stores map an object {\emph key} to one or more physical storage nodes (if replication is used). Depending on the implementation, a request for a key might always go to the same physical node or load balanced across all replicas. As detailed in Section~\ref{sec:measurement}, one has the option of using unique keys for each chunk of an object or share the same key across chunks but assign them different byte ranges. The former wastes significant storage capacity, whereas the latter will likely demonstrate higher correlation across parallel reads/writes of distinct chunks of the same object. Nonetheless, our measurements in different regions over a popular public cloud establish that in fact sharing the same key results in reasonably well weak-correlations enabling parallelism and coding gains. However, our measurements also indicate that indeed universally good performance is not guaranteed as one region fails to deliver this weak-correlation.

\item Exact analysis for computing the optimal code rate and chunking level is far beyond trivial. In Sections~\ref{ssec:ana:formulation} to \ref{ssec:ana:optStatic}, we relate the delay optimal values for chunking level and code rate to the queue backlogs via an approximate queuing analysis.

\item Using this analysis, in Section~\ref{ssec:ana:adaptive},  we introduce  \ourproposal (Throughput Optimal FEC Cloud) that implements dynamic adjustment of chunking and redundancy levels to provide the optimal throughput-delay trade-off. In other words, \ourproposal achieves the lower envelope of curves in all colors in Fig.\ref{fig:fixedDelays}.

\item The primary novelty of \ourproposal is in its backlog-based adaptive algorithm for dynamically adjusting the chunk size as well as the number of redundant requests issued to fulfill storage access requests. This algorithm of variable chunk sizing can be viewed as a novel integration of prior observations from the two bodies of works discussed above. Based on the observed backlog level as an indicator of the workload, \ourproposal increases or decreases the chunk size, as well as the number of redundant requests. In our trace-driven evaluations, we demonstrate that: (1) \ourproposal successfully adapts to full range of workloads, delivering $3\times$ lower average delay than the basic static strategy without chunking under light workloads, and under heavy workloads over $3\times$ the throughput of a static strategy with high chunking and redundancy levels optimized for service delay; and (2) \ourproposal provides good QoS guarantees as it delivers low delay variations.

\comment{
\ourproposal works without any explicit information from the back-end cloud storage implementation: its adaptation strategy is implemented solely at the front-end application server (the storage client) and is based exclusively on the measured latency from unmodified cloud storage systems. This allows \ourproposal to be more easily deployed, as individual cloud applications can adopt \ourproposal without being tied-up with any particular cloud storage system, as long as a small number of APIs are provided by 
most existing popular cloud storage systems such as Amazon S3 and Microsoft's Azure Storage.
}

\item Although \ourproposal does not need any explicit information about the internal operations of the storage cloud, it needs to log latency performance and model the cumulative distribution of the delay performance of the storage cloud. We also propose a greedy heuristic that does not need to build such a model, and via trace-driven simulations we show that its performance on average latency is on a par with the performance of \ourproposal, but exhibiting significantly higher performance variations.

\end{itemize}

\section{System Models}
\label{sec:system}

\subsection{Basic Architecture and Functionality}
The basic system architecture captures how Internet services today utilize public or private storage clouds.  The architecture consists of proxy servers in the front-end and a key-value store, referred to as  storage cloud, in the back-end. 
Users interact with the proxy through a high-level API and/or user interfaces.  
The proxy translates every high-level user request (to read or write a file) into a set of $n \ge 1$ tasks. Each task is essentially a basic storage access operation such as {\tt put, get, delete,} etc. that will be accomplished using low-level APIs provided by the storage cloud. The proxy maintains a certain number of parallel connections to the storage cloud and each task is executed over one of these connections. After a certain number of tasks are completed successfully, the user request is considered accomplished and the proxy responds to the user with an acknowledgment. The solutions we present are deployed on the proxy server side transparent to the storage cloud.

For read request, we assume that the file is pre-coded into $n^{max}\ge n$ coded chunks with an $(n^{max},k)$ MDS code and stored on the cloud. Completion of downloading any $k$ coded chunks provides sufficient data to reconstruct the requested file. Thus, the proxy decodes the requested file from the $k$ downloaded chunks and replies to the client. The $n-k$ unfinished and/or not-yet-started tasks are then canceled and removed from the system. 

\begin{figure}[!t]
\centering
\includegraphics[width = \columnwidth]{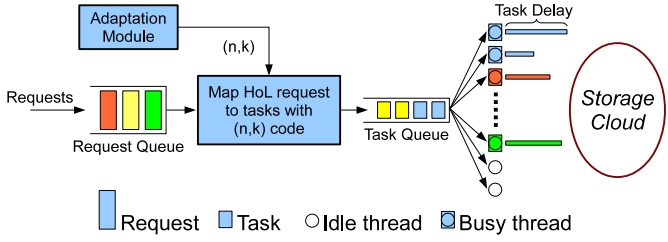}
\caption{System Model}
\label{fig:system}
\end{figure}

For write request, the file to be uploaded is divided and encoded into $n$ coded chunks using an $(n,k)$ MDS code and hence completion of uploading any $k$ coded chunks means sufficient data have been stored onto the cloud. 
Thus, upon receiving $k$ successful responses from the storage cloud, the proxy sends a {\em speculative} success response to the client, without waiting for the remaining $(n-k)$ tasks to finish.
Such speculative execution is a commonly practiced optimization technique to reduce client perceived delay in many computer systems such as databases and replicated state machines \cite{zyzzyva}.
Depending on the subsequent read profile on the same file, the proxy can (1) continue serving the remaining tasks till all $n$ tasks finish, or (2) change them to low priority jobs that will be served only when system utilization is low, or (3) cancel them preemptively. The proxy can even (4) run a demon program in the background that generates all $n_{max}$ coded chunks from the already uploaded chunks when the system is not busy.

Accordingly, we model the proxy by the queueing system shown in Fig.\ref{fig:system}. There are two FIFO (first-in-first-out) queues: (i) the {\em request queue} that buffers all incoming user requests, and (ii) the {\em task queue} that is a multi-server queue and holds all tasks waiting to be executed. $L$ threads\footnote{We avoid the term ``server'' that is commonly used in queueing theory literature to prevent confusion.}, representing the set of parallel connections to the storage cloud, are attached to the task queue. 
The adaptation module of \ourproposal monitors the state of the queues and the threads, and decides what coding parameter $(n,k)$ to be used for each request. 
Without loss of generality, we assume that the head-of-line (HoL) request leaves the request queue only when there is at least one idle thread {\bf and} the task queue is empty. A batch of $n$ tasks are then created for that request and injected into the task queue. As soon as any $k$ tasks complete successfully, the request is considered completed.  Such a queue system is work conserving since no thread is left idle as long as there is any request or task pending.

\begin{figure}[t]
\centering
\includegraphics[width = \columnwidth]{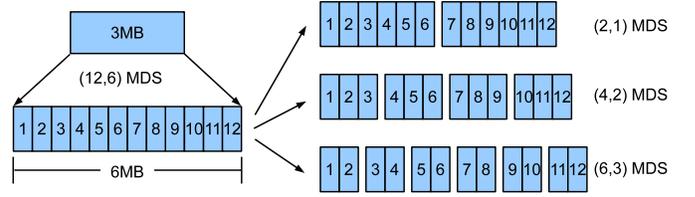}
\caption{Example of supporting multiple chunk sizes with Shared Key approach: the 3MB file is divided and encoded into a coded file of 6MB consisting 12 strips, each of 0.5MB. Download the file using a $(2,1)$ MDS code is accomplished by creating two read tasks: one for strips 1-6, and the other for strips 7-12.} 
\label{fig:partialRead}
\end{figure}

\subsection{Basics of Erasure Codes}
\label{ssec:model:code}
An $(n,k)$ MDS code (e.g., Reed-Soloman codes) encodes $k$ data chunks each of $B$ bits into a codeword consisting of $n$ $B$-bit long coded chunks. The coded chunks can sustain up to $(n-k)$ erasures such that the $k$ original data chunks can be efficiently reconstructed from {\bf any} subset of $k$ coded chunks.
$n$ and $k$ are called the length and dimension of the MDS code. We also define $r = n/k$ as the redundancy ratio of an $(n,k)$ MDS code. The erasure resistant property of MDS codes has been utilized in prior works \cite{fastcloud,Longbocodeingincloud, MDS-queue}, as well as in this paper, to improve delay of cloud storage systems. Essentially a coded chunk experiencing long delay is treated as an erasure.

In this paper, we make use of another interesting property of MDS codes to implement variable chunk sizing of \ourproposal in a storage efficient manner:  MDS code of high length and dimension for small chunk size can be used as MDS code of smaller code length and dimension for larger chunk size. To be more specific, consider any $(N,K)$ MDS code for chunks of $b$ bits. To avoid confusion, we will refer to these $b$-bit chunks as strips. A different MDS code of length $n = N/m$, dimension $k=K/m$ and chunk size $B=b m$ for some $m>1$ can be constructed by simply batching every $m$ data/coded strips into one data/coded chunk. The resulting code is an $(n,k)$ MDS code for $B$-bit chunks because any $k$ coded chunks covers $mk = K$ coded strips, which is sufficient to reconstruct the original file of $Bk = b m \times K/m = bK$ bits.  This property is illustrated as an example in Fig. \ref{fig:partialRead}. In this example, a 3MB file is divided into 6 strips of 0.5MB and encoded into 12 coded strips of total size 6MB, using a $(12,6)$ MDS code. This code can then be used as a $(2,1)$ code for 3MB chunks, a $(4,2)$ code for 1.5MB chunks and a $(6,3)$ code for 1MB chunks {\bf simultaneously} by batching 6, 3 and 2 strips into a chunk.

\subsection{Definitions of Different Delays}
The delay experienced by a user request consists of two components: {\em queueing delay ($D_q$)} and {\em service delay ($D_s$)}. Both are defined with respect to the request queue: (i) the queueing delay is the amount of time a request spends waiting in the request queue and (ii) the service delay is the period of time between when the request leaves the request queue (i.e., admitted into the task queue and started being served by at least one thread) and when it finally leaves the system (i.e., the first time when any $k$ of the corresponding tasks complete). In addition, we also consider the {\em task delays ($D_t$)}, which is the time it takes for a thread to serve a task assuming it is not terminated or canceled preemptively. To clarify these definitions of delays, consider a request served with an $(n,k)$ MDS code, with $T_A$ being its arrival time, $T_1\le T_2\le \cdots \le T_n$ the starting times of the corresponding $n$ tasks\footnote{We assume $T_i = \infty$ if the $i$-th task is never started.}. Then the queueing delay is $D_q = T_1 - T_A$. Suppose $D_{t,1},\cdots,D_{t,n}$ are the corresponding task delays, then the completion times of these task will be $X=\{T_1+D_{t,1},\cdots,T_n+D_{t,n}\}$ if none is canceled. So the request will leave the system at time $X_{(k)}$, which denotes the $k$-th smallest value in $X$, i.e., the time when $k$ tasks complete. Then the service delay of this request is $D_s = X_{(k)} - T_1$.

\section{Variable Chunk Sizing}
\label{sec:measurement}

\ifnewtext
In this section, we discuss implementation issues as well as pros and cons of two potential approaches, namely {\em Unique Key} and {\em Shared Key}, for supporting erasure-code-based access to files on the storage cloud with a variety of chunk sizes. Suppose the maximum desired redundancy ratio is $r$, then these approaches implement variable chunk sizing as follows:
\begin{itemize}
\item {\bf Unique Key:} For every choice of chunk size (or equivalently $k$), a separate batch of $rk$ coded chunks are created and each coded chunk is stored as an individual object with its unique key on the storage cloud. The access to different chunks is implemented through basic {\tt get, put} storage cloud APIs. 

\item {\bf Shared Key:} A coded file is first obtained by stacking together the coded strips obtained by applying a high-dimension $(N=rK,K)$ MDS code to the original file, as described in Section \ref{ssec:model:code} and illustrated in Fig.\ref{fig:partialRead}. 
For read, the coded file is stored on the cloud as one object. Access to chunks with variable size is realized by downloading segments in the coded file corresponding to batches of a corresponding number of strips, using the same key with more advanced ``partial read'' storage cloud APIs. Similarly, for write, the file is uploaded in parts using ``partial write'' APIs and then later merged into one object in the cloud.
\end{itemize}

\subsection{Implementation and Comparison of the two Approaches}
\label{ssec:measurement:partialRead}

\subsubsection{Storage cost} When the user request is to write a file, storage cost of Unique Key and Shared Key is not so different. However, to support variable chunk sizing for read requests, Shared Key is significantly more cost-efficient than Unique Key. With Shared Key, a single coded file stored on the cloud can be reused to support essentially an arbitrary number of different chunk sizes, as long as the strip size is small enough. On the other hand, it seems impossible to achieve similar reusing with the Unique Key approach where different chunks of the same file is treated as individual objects. So with Unique Key, every additional chunk size to be supported requires an extra storage cost $r\times$ file size. Such linear growth of storage cost easily makes it prohibitively expensive even to support a small number of chunk sizes.

\subsubsection{Diversity in delays} The success of \ourproposal and other 
proposals to use redundant requests (either with erasure coding or replication) for delay improvement relies on diversity in cloud storage access delays. In particular, \ourproposal, as well as \cite{fastcloud,Longbocodeingincloud,MDS-queue}, requires access delays for different chunks of {\bf the same file} to be weakly correlated.
  
With Unique Key, since different chunks are treated as individual objects, there is no inherent connection among them from the storage cloud system's perspective. So depending on the internal implementation of object placement policy of the storage cloud system, chunks of a file can be stored on the cloud in different storage units (disks or servers) on the same rack, or in different racks in the same data center, or even to different data centers at distant geographical locations. Hence it is quite likely that delays for accessing different chunks of the same file show very weak correlation. 

On the other hand, with Shared Key, since coded chunks are combined into one coded file and stored as one object in the cloud, it is very likely that the whole coded file, hence all coded chunks/strips, is stored in the same storage unit, unless the storage cloud system internally divides the coded file into pieces and distributes them to different units. Although many distributed storage systems do divide files into parts and store them separately, it is normally only for larger files. For example, the popular Hadoop distributed file system by default does not divide files smaller than 64MB. When different chunks are stored on the same storage unit, we can expect higher correlation in their access delays. It then is to be verified that the correlation between different chunks with the Shared Key approach is still weak enough for our coding solution to be beneficial.

\begin{figure*}[!t]
\centering
	\subfigure[Region US Standard]{
		\label{fig:ccdf:thread:virginia1}
		\includegraphics[width=\threewidth]{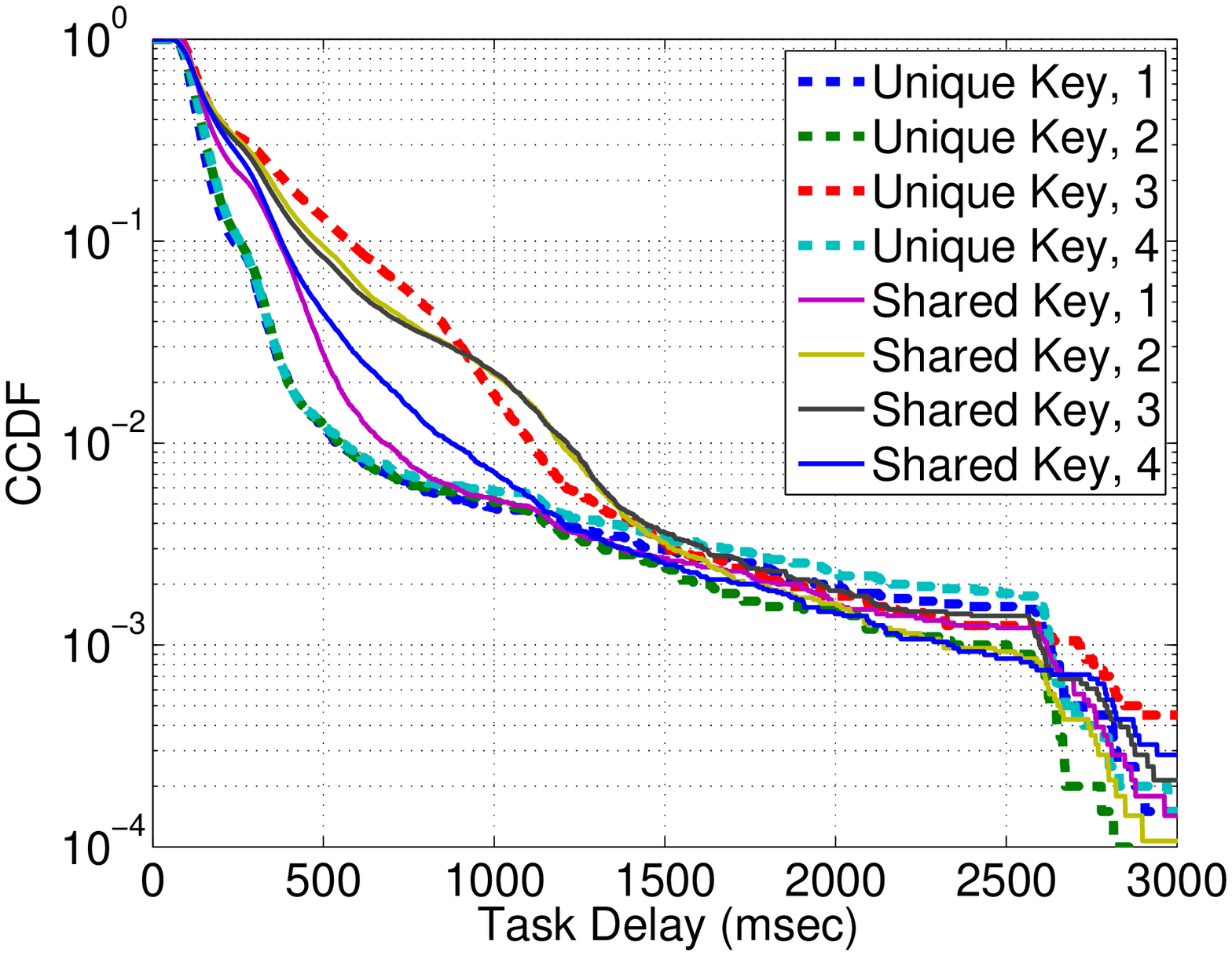}
	}
\hfill
	\subfigure[Region US Standard at a different time]{
		\label{fig:ccdf:thread:virginia2}
		\includegraphics[width=\threewidth]{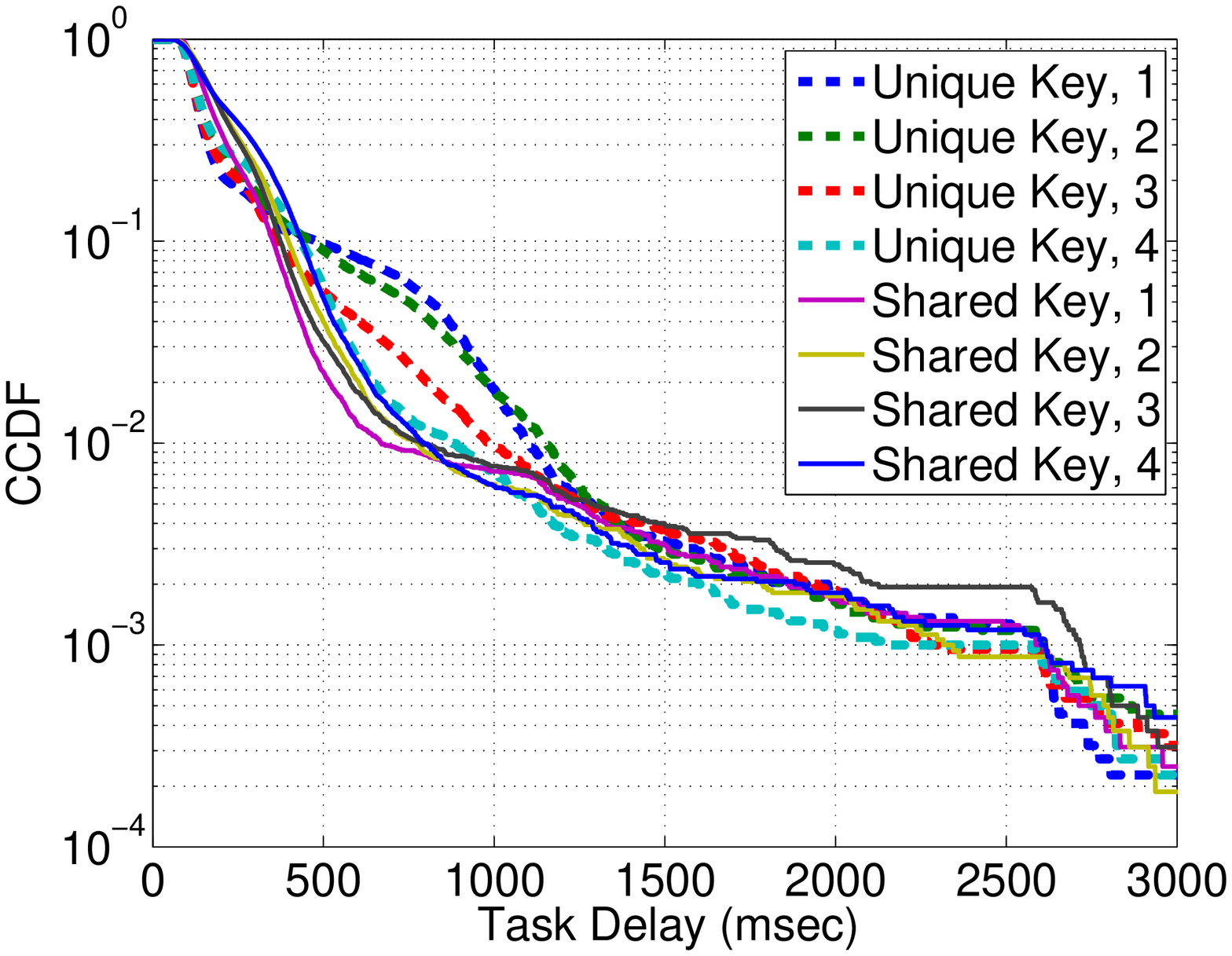}
	}
\hfill
		\subfigure[Region North California]{
			\label{fig:ccdf:thread:northCal}
			\includegraphics[width=\threewidth]{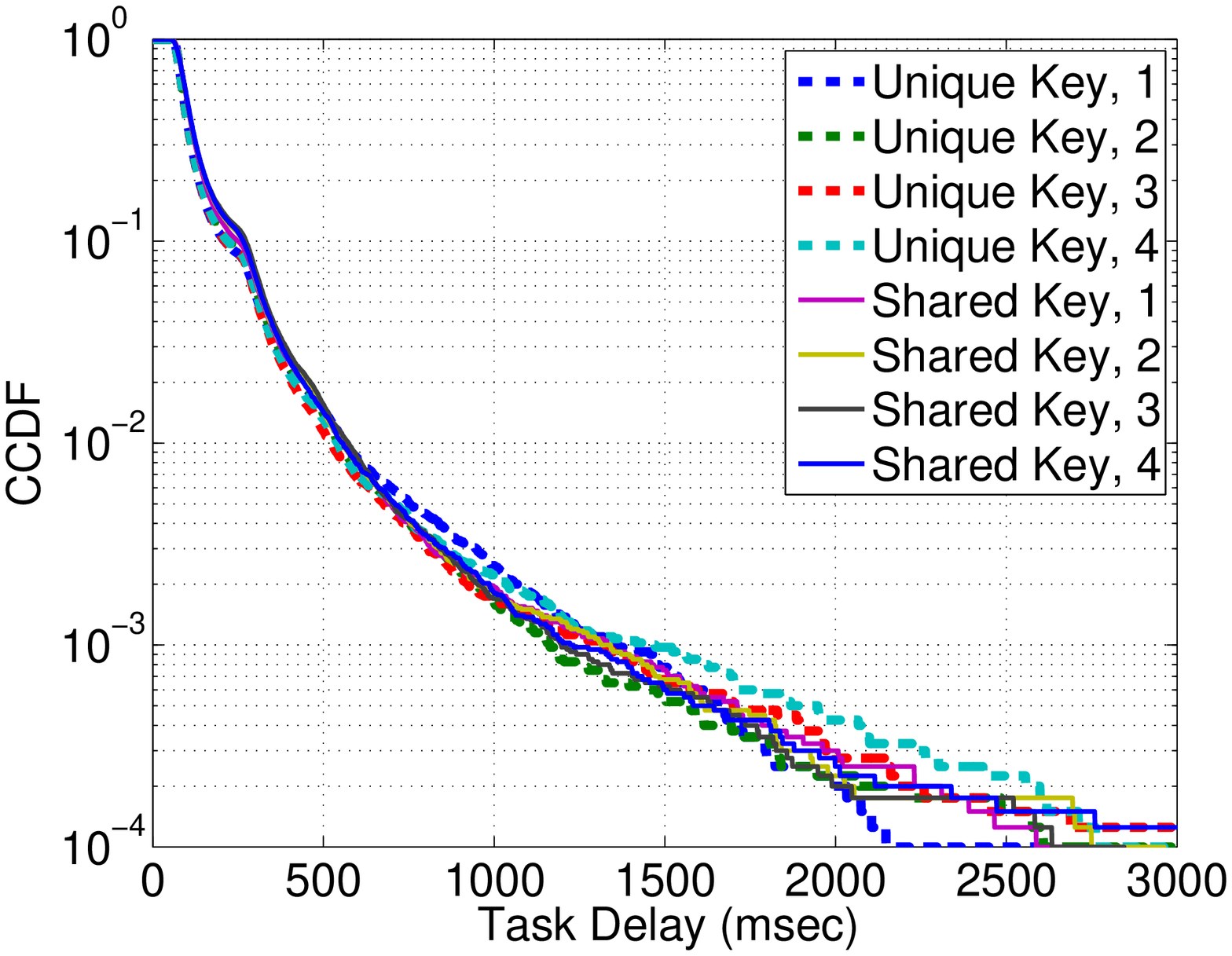}
		}%
\caption{CCDF of individual threads with 1MB chunks and $n=4$, measured on May 1st, 2013}
\label{fig:ccdf:thread}
\end{figure*}

\begin{figure*}[!t]
\centering
\null\hfill
	\subfigure[Unique Key]{
		\label{fig:ccdf:uniqueKey}
		\includegraphics[width=\onewidth ]{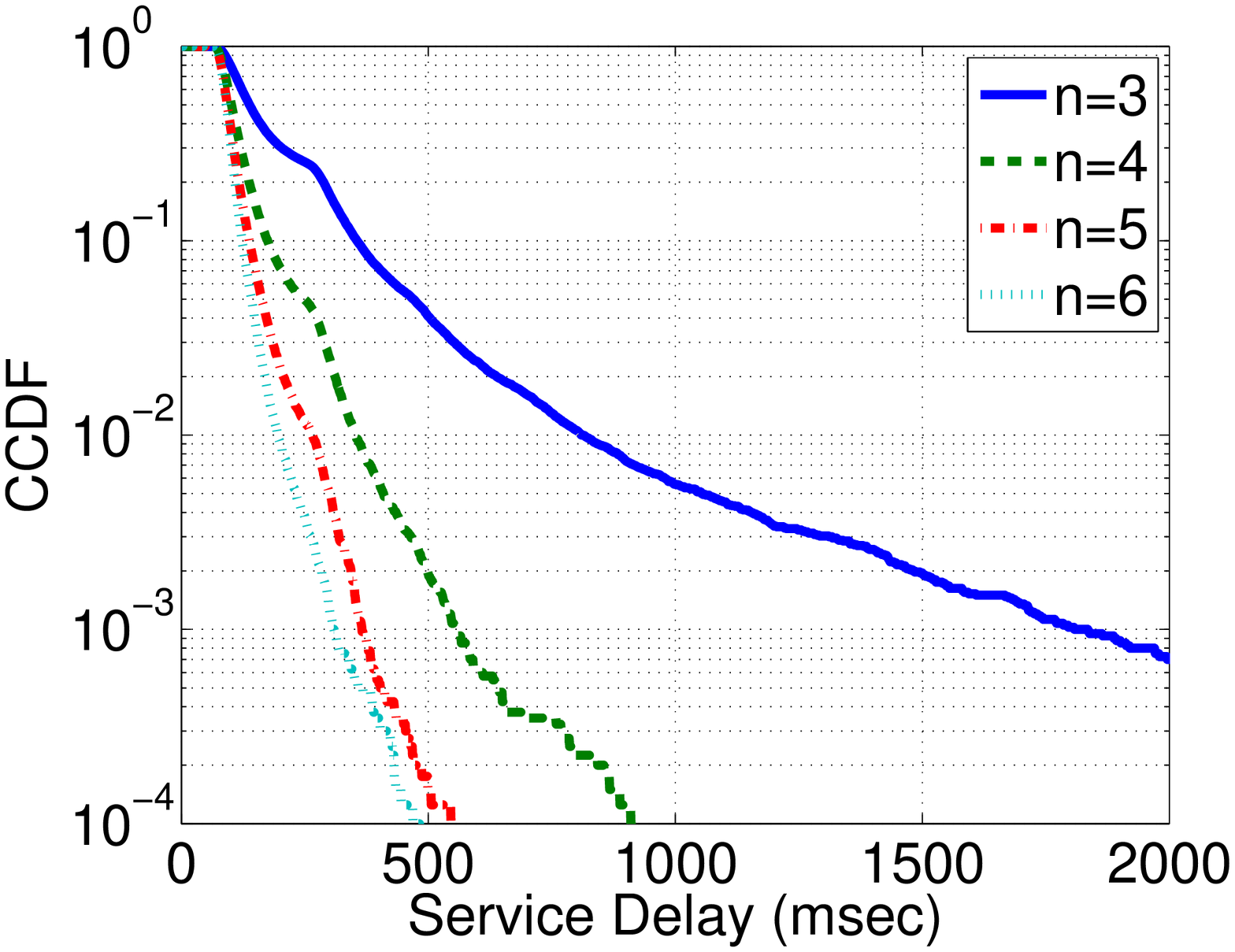}
	}
\hfill
	\subfigure[Shared Key]{
		\label{fig:ccdf:partialRead}
		\includegraphics[width=\onewidth ]{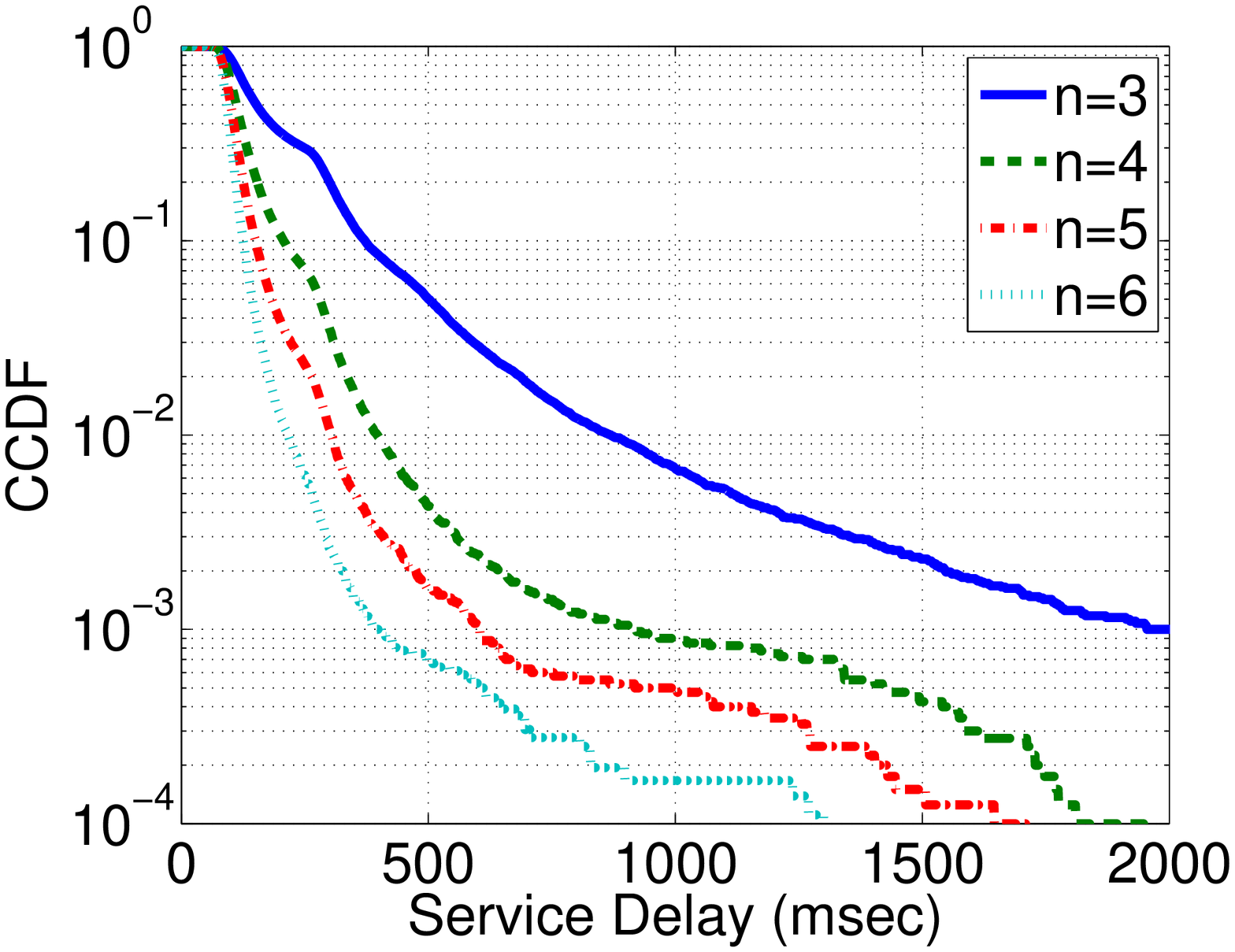}
	}
	\hfill\null
\caption{CCDF of service delay for reading 3MB files with 1MB chunks}
\label{fig:CCDF:FEC}
\end{figure*}

\subsubsection{Universal support}
Unique Key is the approach adopted in our previous work \cite{fastcloud} to support erasure-code based file accessing with {\bf one predetermined} chunk size. A benefit of Unique Key is that it only requires basic {\tt get} and {\tt put} APIs that all storage cloud systems must provide. So it is readily supported by all storage cloud systems and can be implemented on top of any one.

On the other hand, Shared Key requires more advanced APIs that allow the proxy to download or upload only the targeted segment of an object. Such advanced APIs are not currently supported by all storage cloud systems. For example, to the best of our knowledge currently Microsoft's Azure Storage provides only methods for  ``partial read''\footnote{E.g. {\tt DownloadRangeToStream(target, offset, length)} downloads a segment of {\tt length} bytes starting from the {\tt offset}-th byte of the {\tt target} object (or ``blob'' in Azure's jargon).}
 but none for ``partial write''. On the contrary, Amazon S3 provides partial access for both read and write: the proxy can download a specific inclusive byte range within an object stored on S3 by calling {\tt getObject(request,destination)}\footnote{The byte range is set by calling {\tt request.setRange(start,end)}.}; and for uploading an {\tt uploadPart} method to upload segments of an object and an {\tt completeMultipartUpload} method to merge the uploaded segments are provided. We expect more service providers to introduce both partial read and write APIs in the near future.

\comment{

Nowadays, almost all popular storage cloud systems implement their own internal object replication mechanisms for very high availability and durability. These internal replication mechanisms usually try to place copies of an object to different storage units (disks or servers) in the same rack, different racks in the same data center, or even different data centers at distant geographical locations so that it is very unlikely that different copies will face the same network bottleneck and/or lost due to unexpected disaster. As a results,  

}

%

%
\comment{
In FAST CLOUD, the chunk size for each file is assumed to be predetermined and fixed, so it is plausible and cost-effective to treat each coded chunk as a individual object. However, this approach (referred as ``unique key'' hereafter) will easily be too expensive if more than one chunk sizes are to be supported because its storage cost grows linearly to the number of chunk sizes. For example, suppose we want to support downloading of a 3MB file using 3 chunk sizes of 3MB (no chunking), 1.5MB and 1MB ($k = 1, 2, 3$) and a maximum redundancy ratio of 2 ($n\le 2k$). One needs to create 2 objects of 3MB, 4 objects of 1.5MB and 6 objects of 1MB, resulting in a total storage cost of 18MB which is $6\times$ of the original file size instead of the desired redundancy ratio of $2$.
}

\fi

\subsection{Measurements on Amazon S3}
\label{ssec:measurement:S3}

\begin{figure*}[!t]
\centering
\null\hfill
	\subfigure[Mean]{
		\label{fig:meanTaskDelay}
		\includegraphics[width= \onewidth ]{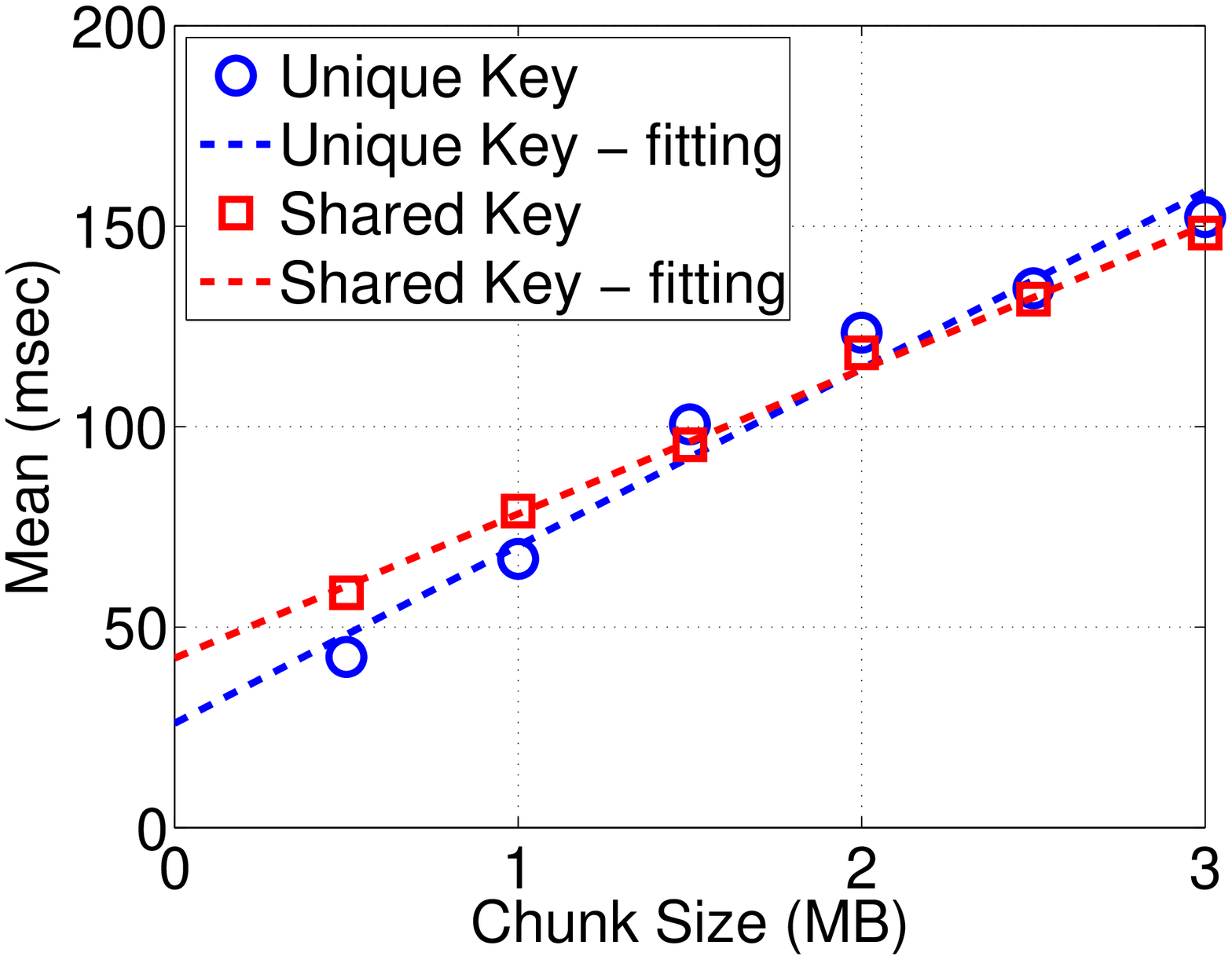}
	}%
\hfill
	\subfigure[Standard Deviation]{
		\label{fig:stdTaskDelay}
		\includegraphics[width= \onewidth ]{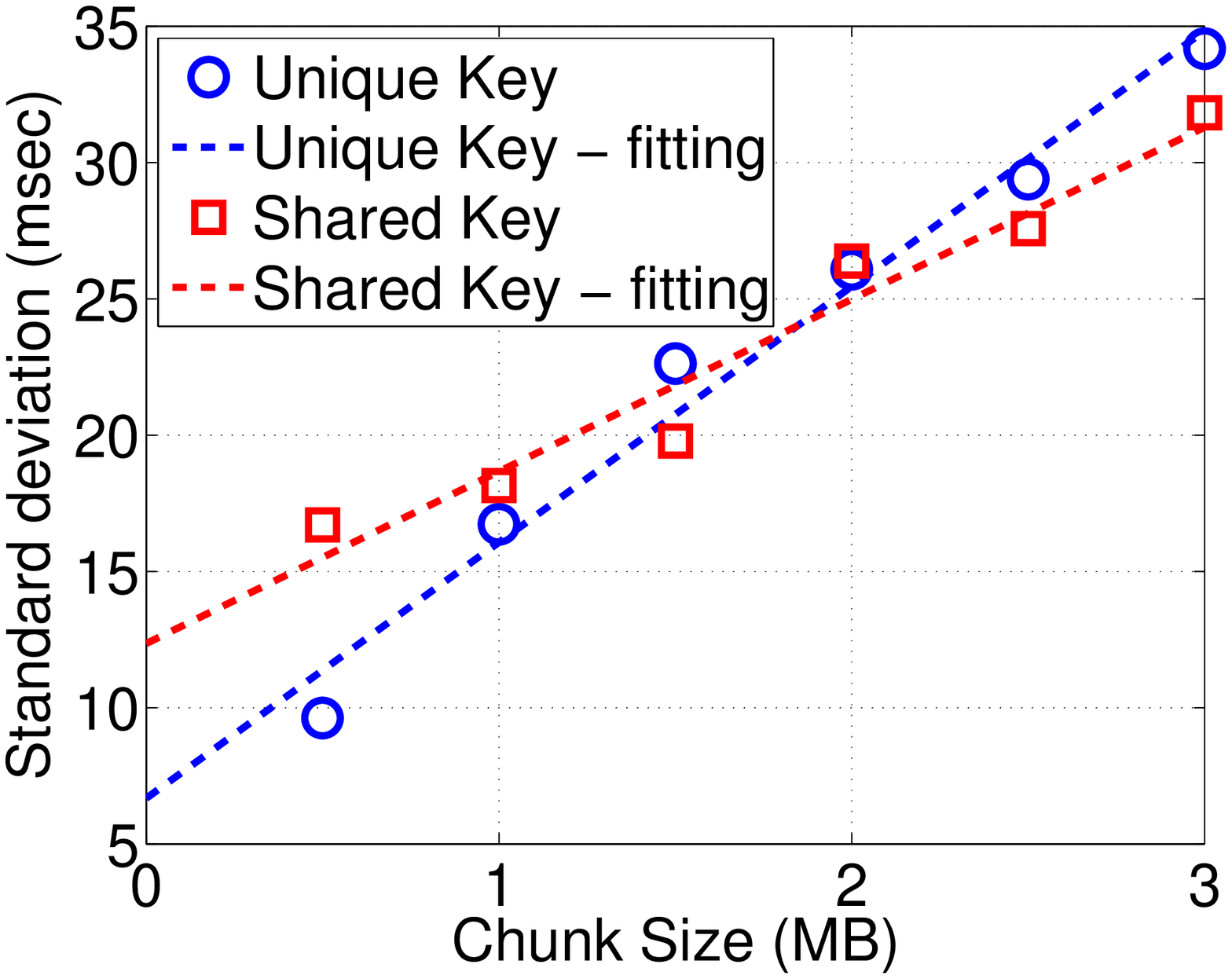}
	}
\hfill\null
\caption{Delay statistics vs. chunk size}
\label{fig:delay-chunkSize}
\end{figure*}

%
To understand the trade-off between Unique Key and Shared Key, we run measurements over Amazon EC2 and S3. EC2 instance served as the proxy in our system model. We instantiated an extra large EC2 instance with high I/O capability in the same availability region as the S3 bucket that stores our objects. We conducted experiments on different week days in May to July 2013 with various chunk sizes between 0.5MB to 3MB and up to $n=12$ coded chunks per file. For each value of $n$, we allow $L=n$ simultaneously active threads while the $i$-th thread being responsible for downloading the $i$-th coded chunk of each file. Each experiment lasted longer than 24 hours. We alternated between different settings to capture similar time of day characteristics across all settings. 

The experiments are conducted within all 8  availability regions in Amazon S3. 
Except for the ``US Standard'' availability region, all other 7 regions demonstrate similar performance statistics that are consistent over different times and days. On the other hand, the performance of ``US Standard'' demonstrated significant variation even at different times in the same day, as illustrated in Fig.\ref{fig:ccdf:thread:virginia1} and Fig.\ref{fig:ccdf:thread:virginia2}. We conjecture that the different and inconsistent behavior of ``US Standard'' might be due to the fact that it targets a slightly different usage pattern and it may employ a different implementation for that reason\footnote{See \url{http://docs.aws.amazon.com/general/latest/gr/rande.html#s3_region}}. We will exclude ``US Standard'' from subsequent discussions. 
For conciseness, we only show a limited subset of findings for availability region ``North California'' that are representative for regions other than ``US Standard'':

(1)
\label{obs:identicalDistribution} 
In both Unique Key and Shared Key, the task delay distribution observed by different threads are almost identical. The two approaches are indistinguishable  even beyond 99.9th percentile. Fig.\ref{fig:ccdf:thread:northCal} shows the complementary cumulative distribution function (CCDF) of task delays observed by individual threads for 1MB chunks and $n=4$. Both approaches demonstrate large delay spread in all regions.

(2)
\label{obs:weakCorrelation}
Task delays for different threads in Unique Key show close to zero correlation, while they demonstrate slightly higher correlation in Shared Key, as it is  expected.
With all different settings, the cross correlation coefficient between different threads stays below 0.05 in Unique Key and ranges from 0.11 to 0.17 in Shared Key. 
Both approaches achieve significant service delay improvements. 
Fig.\ref{fig:CCDF:FEC} plots the CCDF of service delays for downloading 3MB files with 1MB chunks ($k=3$) with $n= 3 \sim 6$, assuming all $n$ tasks in a batch start at the same time. In this setting, both approaches reduce the 99th percentile delays by roughly 50\%, 65\% and 80\% by downloading 1, 2 and 3 extra coded chunks. Although Shared Key demonstrates up to 3 times higher cross correlation coefficient, there is no meaningful statistical distinction in service delay between the two approaches until beyond the 99th percentile.
All availability regions experience different degrees of degradation at high percentiles with Shared Key due to the higher correlation. 
Significant degradation emerges from around 99.9th percentile and beyond in all regions except for ``Sao Paulo'', in which degradation appears around 99th percentile.

(3)
\label{obs:lowerBound} 
Task delays are always lower bounded by some constant $\Delta\ge 0$ that grows roughly linearly as chunk size increases. This constant part of delay cannot be reduced by using more threads: see the flat segment at the beginning of the CCDF curves in Fig.\ref{fig:ccdf:thread} and Fig.\ref{fig:CCDF:FEC}. Since this constant portion of task delays is unavoidable, it leads to the negative effect of using larger $n$ since there is a minimum cost of system resource of $n\Delta $ (time$\times$thread) that grows linearly in $n$. This cost leads to a reduced capacity region for using more redundant tasks, as illustrated in the example of Fig.\ref{fig:fixedDelays}. 
We observe that the two approaches deliver almost identical total delays (queueing + service) for all arrival rates, in spite of the degraded service delay with Shared Key at very high percentile. So we only plot the results with Shared Key in Fig.\ref{fig:fixedDelays}.

(4)
\label{obs:linearGroth}
Both the mean and standard deviation of task delays grow roughly linearly as chunk size increases. Fig.\ref{fig:delay-chunkSize} plots the measured mean and standard deviation of task delays in both approaches at different chunk sizes. Also plotted in the figures are least squares fitted lines for the measurement results. As the figures show, performance of Unique Key and Shared Key are comparable also in terms of how delay statistics scale as functions of the chunk size. Notice that the extrapolations at chunk size = 0 are all greater than zero. We believe this observation reflects the costs of non-I/O-related operations in the storage cloud that do not scale proportionally to object size: for example, the cost to locate the requested object. We also believe such costs contribute partially to the minimum task delay constant $\Delta$.

~

\noindent
{\bf SUMMARY:}
Our measurement study shows that dynamic chunking while preserving weak correlation across different chunks is realizable through both Unique Key and Shared Key. We believe Shared Key is a reasonable choice for implementing dynamic chunking given that it is able to deliver delay performance comparable to Unique Key at a much lower cost of storage capacity. We turn our attention on how to pick the best choices of chunking and FEC rate in the remaining parts of the paper. 

\subsection{Model of Task Delays}
\label{ssec:model:delay}
For the analysis present in the next section,
we model the task delays as independently distributed random variables whose mean and standard deviation grow linearly as chunk size $B$ increases. More specifically, we assume the task delay $D_t$ for chunk size $B$ following distribution in the form of 
\begin{equation}
D_t(B) \sim \Delta(B) + exp(\mu(B)),
\end{equation}
where 
$
\Delta(B) = \fixedDelta + \linearDelta B
$
captures the lower bound of task delays as in observation (3), and $exp(\mu(B))$ represents a exponential random variable that models the tail of the CCDF. The mean and standard deviation of the exponential tail both equal to  
$
\frac{1}{\mu(B)} = \fixedExp + \linearExp B.
$
With this model, constants
$\fixedDelta$ and $\fixedExp$ together capture the non-zero extrapolations of the mean and standard deviation of task delays at chunk size 0, and similarly, constants $\linearDelta$ and $\linearExp$ together capture the rate at which the mean and standard deviation grow as chunk size increases, as in observation (4).

\comment{
We assume there are $m\ge 1$ classes of requests. Requests of each class have identical file size  and all are divided into chunks of identical size. Under this assumption, service times of all chunks of the same class follow the same distribution and each class $i$ can be characterized by a three-tuple $(k_i,\Delta_i,\mu_i)$, where $\Delta_i$ and $\mu_i$ specifies the delay distribution of class-$i$ chunks.
Throughout this paper, we assume $k_i$'s (and accordingly chunk sizes) are determined a priori and $(\Delta_i,\mu_i)$ are given. Our focus will be on the adaptation/choice of $n_i$'s.  
}

\section{Design of \ourproposal}
\label{sec:proposed-algorithm}

For the analysis in this section, we group requests into classes according to the tuple {\tt (type, size)}. Here {\tt type} can be read or write, and can potentially be other type of operations supported by the cloud storage. Each type of operation has its own set of delay parameters $\{\fixedDelta, \linearDelta, \fixedExp, \linearExp\}$. Subscripts will be used to indicate variables associated with each class. We use $n_i$, $k_i$ and $r_i$ to denote the code length, dimension and redundancy ratio for the code used to serve class $i$ requests. Also let $p_i$ denote the fraction of total arrivals contributed by class $i$. We use vectors $\lenVec$, $\dimVec$, $\rateVec$ and $\compVec$ to denote the collection of corresponding variables for all classes. 

The system throughput is defined as the average number of successfully served requests per unit time. The {\bf static code capacity} $\CapSta(\compVec,\dimVec,\rateVec)$  is defined as the maximum deliverable throughput, assuming $\compVec$, $\dimVec$, and $\rateVec$ are fixed. The {\bf full capacity} $\CapFull(\compVec)$ is then defined as the maximum static code capacity considering all possible choices of $(\dimVec,\rateVec)$ with $\compVec$ fixed. For a given request arrivals rate $\lambda$, the system throughput equals to the smaller of $\lambda$ and the (static or full) capacity.

\subsection{Problem Formulation and Main Result for Static Strategy}
\label{ssec:ana:formulation}
Given total arrival rate $\lambda$ and composition of requests $\compVec$, we want to find the best choice of FEC code for each class such that the average expected total delay is minimized. Relaxing the requirement for $n_i$ and $k_i$ being integers, this is formulated as the following minimization problem\footnote{Notice that all classes share the same queueing delay. Also, we require $k_i\ge 0$ instead of $k_i\ge 1$ for a technicality to simplify the proof of the uniqueness of the optimal solution. We require $r_i\ge 1$ since $n_i\ge k_i$. $\lambda < \CapSta(\compVec,\dimVec,\rateVec)$ is imposed for queue stability.}:
\begin{align*}
\min_{\dimVec,\rateVec} &~~~ D_q + \sum_i p_i D_{s,i} 
\label{eq:optimization}
\tag{$\ast$}
\\
\mathrm{s.t.} &~~~ k_i\ge 0,~~ r_i \ge 1~~\forall i,\\
&~~~ \lambda < \CapSta(\compVec,\dimVec,\rateVec). 
\end{align*}
In the above formulation, we use $\dimVec$ and $\rateVec$ as the optimizing variables, instead of a more intuitive choice of $\lenVec$ and $\dimVec$. This choice helps simplify the analysis because $\dimVec$ and $\rateVec$ can be treated as independent variables while $\lenVec$ being subject to the constraint $\lenVec\ge \dimVec$.
In  subsequent sections, we first introduce approximations for the expected queueing and service delays assuming that the FEC code used to serve requests of each class is predetermined and fixed (Section \ref{ssec:ana:static}).
Then we show that optimal solutions to the above non-convex optimization problem exhibit the following property (Section \ref{ssec:ana:optStatic}):

\mybox{
The optimal values of $n_i$, $k_i$ and $r_i$ can all be expressed as functions solely determined by $Q$ -- the expected length of the request queue:
$$n_i = N_i(Q),~~k_i = K_i(Q) \text{~~and~~}  r_i = R_i(Q).$$
$N_i$, $K_i$ and $R_i$ are all strictly decreasing functions of $Q$.
}
%

~

\noindent This finding is then used as the guideline in the design of our backlog-driven adaptive strategy \ourproposal (Section \ref{ssec:ana:adaptive}).

\subsection{Approximated Analysis of Static Strategies}
\label{ssec:ana:static}
\begin{figure}[!t]
\centering
\includegraphics[width= \onewidth]{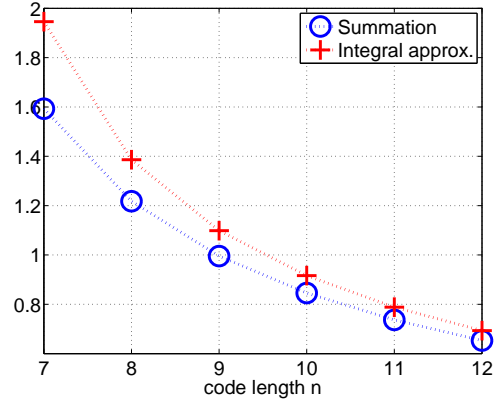}
\caption{Comparison of the summation term in Eq.\ref{eq:serviceDelay} and integral approximation for $k=6$.}
\label{fig:approxGap}
\end{figure}

Denote $J_i$ as the file size of class $i$.
Consider a request of class $i$ served with an $(n_i,k_i)$ MDS code, i.e., $B_i = J_i/k_i$. First suppose {\em all $n_i$ tasks start at the same time}, i.e., $T_1 = T_{n_i}$. In this case, given our model for task delays, it is trivial to show that the expected service delay equals to 
\begin{align}
D_{s,i} =& \Delta_i(J_i/k_i)
				+ \frac{1}{\mu_i(J_i/k_i)}
				  \sum_{j=n_i-k_i+1}^{n_i}\frac{1}{j}
\nonumber\\
\approxeq& \Delta_i(J_i/k_i) 
				+ \frac{1}{\mu_i(J_i/k_i)}\ln \left(\frac{n_i}{n_i-k_i} \right)
\nonumber\\
=& \fixedDelta_i + \frac{\linearDelta_i J_i}{k_i} + 
\left(\fixedExp_i + \frac{\linearExp_i J_i}{k_i}\right)\ln \left(\frac{r_i}{r_i-1}\right).
\label{eq:serviceDelay}
\end{align}
For the analysis, we approximate the summation $\sum_{j=n_i-k_i+1}^{n_i}1/j$ with its integral upper bound $\int_{n_i-k_i}^{n_i}\frac{1}{x} dx = \ln \left(\frac{n_i}{n_i-k_i}\right)$. The gap of approximation is always upper bounded by the Euler-–Mascheroni constant $\approxeq 0.577$ for any $n_i- k_i \ge 1$ and quickly diminishes to 0 when $n_i$ gets large, as illustrated in Fig.\ref{fig:approxGap}. Although the gap goes to $\infty$ as $n_i-k_i\rightarrow 0$, it does not really matter for the purpose of this paper since any optimal solution with $n_i$ closer to $k_i$ only means we should set $n_i=k_i$.

Also define the system usage (or simply ``cost'') of a request as the sum of the amount of time each of its tasks being served by a thread\footnote{The time a task $j$ being served is $D_{t,j}$ if it completes successfully; $\left(X_{(k)}-T_j\right)$ if it starts but is terminated preemptively; and 0 if it is canceled while waiting in the task queue.}. When all tasks start at the same time, its expected system usage is (see Section IV of \cite{fastcloud} for detailed derivation)
\begin{align}
U_i =& n_i\Delta_i(J_i/k_i) + \frac{k_i}{\mu_i(J_i/k_i)}\nonumber\\
=& \fixedDelta_i k_i r_i + \linearDelta_i J_i r_i 
+ \fixedExp_i k_i + \linearExp_i J_i.
\end{align}
Given that class $i$ contributes to $p_i$ fraction of the total arrivals, the average cost per request is $\aveUsage = \sum_i p_i U_i$. With $L$ simultaneously active threads, the departure rate of the system as well as the request queue is  $L/\aveUsage$ (request/unit time). In light of this observation, we approximate the request queue with an $M/M/1$ queue with service rate $L/\aveUsage$
\footnote{This $M/M/1$ approximation is a special case of the $M/G/1$ approximation used in \cite{fastcloud}: $D_q = \frac{\beta\lambda \aveUsage^2}{2L(L-\lambda \aveUsage)}$, with $\beta = 1$. Our findings in this paper readily generalizes to accommodate the $M/G/1$ approximation.
}.
In other words, the static code capacity for a given $\compVec$ and fixed code choice $(\dimVec,\rateVec)$ is approximated by
\begin{equation}
\CapSta(\compVec,\dimVec,\rateVec) = \frac{L}{\aveUsage}.
\end{equation}
Let
\begin{align}
\normArrival =& \lambda \aveUsage 
\nonumber \\
=& \lambda \sum_i p_i (\fixedDelta_i k_i r_i + \linearDelta_i J_i r_i 
+ \fixedExp_i k_i + \linearExp_i J_i)
\end{align}
represent the arrival rate of system usage imposed by the request arrivals. Then the last inequality constraint of the optimization  problem \eqref{eq:optimization} becomes 
\begin{equation}
\normArrival < L.
\end{equation}

With the $M/M/1$ queue approximation, the queueing delay in the original system at total arrival rate $\lambda$ is approximated by
\begin{eqnarray}
D_q = \frac{1}{L/\aveUsage - \lambda} - \frac{1}{L/\aveUsage}
= \frac{\lambda \aveUsage^2}{L(L-\lambda \aveUsage)}.
\end{eqnarray}

Noticing that given $\compVec$, the (approximated) static coded capacity $\CapSta(\compVec,\dimVec,\rateVec)=L/\aveUsage$ is maximized when $k_i=1$, $r_i=1$, $\forall i$, we approximate the full capacity $\CapFull(\compVec) = \CapSta(\compVec,\ones,\ones)$, where $\ones$ denotes the all-one vector. We acknowledge that the above approximations are quite coarse, especially because tasks of the same batch do not start at the same time in general.  However, remember that the main objective of this paper is to develop a practical solution that can achieve the optimal throughput-delay trade-off. According to the simulation results, these approximations are sufficiently good for this purpose.

\subsection{Optimal Static Strategy}
\label{ssec:ana:optStatic}

Even with the above approximations, the minimization problem \eqref{eq:optimization} is
not a convex optimization problem: the feasible region is not a convex set due to the $k_i r_i$ terms in $\normArrival$. In general, non-convex optimization problems are difficult to solve. Fortunately, we are able to prove the following theorem according to which this non-convex optimization problem can be solved numerically with great  efficiency.

\begin{theorem}
\label{thm:optimal}
The optimal solutions to \eqref{eq:optimization} must satisfy the following equations, regardless of $\lambda$ and $\compVec$.
\begin{align}
k_i &= \KR_i(r_i)
\nonumber\\
&\triangleq\frac{\fixedDelta_i \FR_i - \linearExp_i J_i + \sqrt{(\fixedDelta_i \FR_i - \linearExp_i J_i)^2 + 4\fixedExp_i \linearDelta_i J_i \FR_i }}{2\fixedExp_i},
~~\forall i
\label{eq:opt:k-function-r}
\end{align}
\begin{align}
\frac{L (\fixedExp_i k_i + \linearExp_i J_i)}{k_i r_i (r_i-1) (\fixedDelta_i k_i + \linearDelta_i J_i)}
= \frac{L (\fixedExp_j k_j + \linearExp_j J_j)}{k_j r_j (r_j-1) (\fixedDelta_j k_j + \linearDelta_j J_j)},
\nonumber\\
~~\forall i,j
\label{eq:opt:invariant-i}
\end{align}
where
$
\FR_i = \frac{J_i r_i(r_i-1)}{\fixedDelta_i r_i + \fixedExp_i}\left(\linearDelta_i + \linearExp_i\ln\frac{r_i}{r_i-1}\right).
$
Moreover, when $\lambda$ and $\compVec$ are given, the optimal solution is the unique solution to the above equations and the one below:
\begin{equation}
\left(\frac{L}{L-\normArrival}\right)^2-1
= \frac{L (\fixedExp_i k_i + \linearExp_i J_i)}{k_i r_i (r_i-1) (\fixedDelta_i k_i + \linearDelta_i J_i)},~~\forall i.
\label{eq:opt:normArrival}
\end{equation}
\end{theorem}
\begin{IEEEproof} 
See Appendix.\\
\end{IEEEproof}
The importance of Theorem \ref{thm:optimal} is two-fold:
\begin{enumerate}
\item With $m$ different classes of requests, the seemingly $2m$-dimension optimization problem is in fact 1-dimensional: According to Eq.\ref{eq:opt:k-function-r}, the optimal $k_i$ is fully determined by the optimal $r_i$ (vice versa). Moreover, according to Eq.\ref{eq:opt:invariant-i}, the optimal $r_i$ further fully determines the optimal choices of $k_j$ and $r_j$ for all other $j\neq i$. In other words, the knowledge of the optimal choice of any $r_i$ (or $k_i$) is sufficient to derive the complete optimal choice of $(\dimVec,\rateVec)$.

\item The optimal solution ($n_i$, $k_i$ and $r_i$) is fully determined by $\normArrival$, hence it is {\em virtually} independent of the particular $\lambda$ and $\compVec$: $\lambda$ and $\compVec$ appear in the above equations only in the form of $\normArrival$ in Eq.\ref{eq:opt:normArrival}. So for any two different pairs of $(\lambda,\compVec)$ and $(\lambda',\compVec')$, as long as $\normArrival = \normArrival'$, they share {\bf the same optimal choice of codes}! An implication of this is that the $m$-class optimization problem can be solved by solving a set of $m$ independent single-class subproblems: the $i$-th subproblem solves for the optimal $(k_i,r_i)$ with class-$i$-only arrivals at rate $\lambda_i$ such that $\lambda_i U_i = \normArrival$, because it is equivalent to the $m$-class problem when $\lambda'=\normArrival/U_i$ and $\compVec'$ such that $p_i' = 1$ and $p_j'=0,~ \forall j\neq i$. 
\end{enumerate}
The second observation above is of particular interest to the purpose of this paper. It suggests that adaptation of different classes can be done separately, as if only arrivals are for the class under consideration. This significantly simplifies the design of our adaptive strategy \ourscheme, resulting in great computational efficiency and flexibility.

\subsection{Adaptive Strategy \ourproposal}
\label{ssec:ana:adaptive}

Despite being the mathematical foundation for the design of \ourproposal, Theorem \ref{thm:optimal} at its current formulation is not very useful in practice. This is because the code adaptation is based on the knowledge of the total workload $\lambda$ and  the popularity distribution of different classes $\compVec$ as per Theorem \ref{thm:optimal}. In practice, both quantities usually demonstrate high degree of volatility, making accurate on-the-fly estimation quite difficult and/or unreliable. So in order to achieve effective code adaptation, a more robust system metric that is easy to measure with high accuracy is desirable. 

Observe that the expected length of the request queue is
\begin{equation}
Q = \lambda D_q= \frac{(\lambda \aveUsage)^2}{L(L-\lambda \aveUsage)} = \frac{\normArrival^2}{L(L-\normArrival)},
\label{eq:Q}
\end{equation}
which can be rewritten as
\begin{equation}
\normArrival = \frac{L\left(\sqrt{Q^2+4Q}-Q\right)}{2}.
\label{eq:rate-from-Q}
\end{equation}
It is trivial and intuitive that $Q$ is a strictly increasing function of $\normArrival$ and vice versa. On the other hand, it is not hard to verify that optimal $n_i$, $k_i$ and $r_i$ are all strictly decreasing functions of $\normArrival$ according to Theorem \ref{thm:optimal}. 
Replacing $\normArrival$ with Eq.\ref{eq:rate-from-Q}, we can conclude the following corollary:
\begin{corollary}
\label{corr:func_of_Q}
The optimal values of $n_i$, $k_i$ and $r_i$ can all be expressed as strictly decreasing functions of $Q$:
\begin{equation}
n_i = N_i(Q),~~k_i = K_i(Q) \text{~~and~~}  r_i = R_i(Q).
\label{eq:func_of_Q}
\end{equation}
\end{corollary}

The findings of Corollary \ref{corr:func_of_Q} conform to the following intuition: 
\begin{itemize}
\item At light workload (small $\lambda$), there should be little backlog in the request queue (small $Q$) and the service delay dominates the total delay. In this case, the system is not operating in the capacity-limited regime. So it is beneficial to increase the level of chunking and redundancy (large $k_i$ and $r_i$) to reduce delay.

\item At heavy workload (larger $\lambda$), there will be a large backlog in the request queue (large $Q$) and the queueing delay dominates the total delay. In this case, the system operates in the capacity-limited regime. So it is better to reduce the level of chunking and redundancy (small $k_i$ and $r_i$) to support higher throughput.
\end{itemize}
More importantly, it suggests the sufficiency to choose the FEC code solely based on the length of the request queue -- a very robust and easy to obtain system metric -- instead of less reliable estimations of $\lambda$ and $\compVec$. As will be discussed later, queue length has other advantages over arrival rate in a dynamic setting.

The basic idea of \ourproposal is to choose $n_i = N_i(q)$ and $k_i= K_i(q)$ for a request of class $i$, where $q$ is the queue length upon the arrival of the request. When this is done to all request arrivals to the system, it can be expected the average code lengths (dimensions) and expected queue length $Q$ satisfy Eq.\ref{eq:func_of_Q}, hence the optimal delay is achieved.  In \ourproposal, this is implemented with a threshold based algorithm, which can be performed very efficiently. 
For each class $i$, we first compute the expected queue length given $n_i \in \left\{1,...,n_i^{max}\right\}$ is the optimal code length by
\begin{equation}
Q^{N}_{i,n_i} = N_i^{-1}(n_i).
\end{equation}
Here $n_i^{max}$ is the maximum number of tasks allowed for a class $i$ request. Since $N_i$ is a strictly decreasing function, its inverse $N_i^{-1}$ is a well-defined strictly decreasing function. As a result, we have
$ Q^{N}_{i,1} > Q^{N}_{i,2} > \cdots > Q^{N}_{i,n_i^{max}} >0.$
Note that our goal is to use code length $n$ if the queue length $q$ is around $Q^{N}_{i,n}$, so we want a set of thresholds $\{\nthreshold_{i,n}\}$ such that
\begin{align*} 
\nthreshold_{i,1}> Q^{N}_{i,1} &>\nthreshold_{i,2} > Q^{N}_{i,2} > \cdots \\
\cdots &> \nthreshold_{i,n_i^{max}} > Q^{N}_{i,n_i^{max}} > \nthreshold_{i,n_i^{max}+1}=0,
\end{align*}
and will use $n$ such that $q\in[\nthreshold_{i,n+1},\nthreshold_{i,n})$.
In our current implementation of \ourproposal, we use
$ \nthreshold_{i,n} = \left(Q^{N}_{i,n} + Q^{N}_{i,n-1}\right)/2$
for  $n = 2,\cdots, n_i^{max}$ and $\nthreshold_{i,1} = \infty$.
A set of thresholds $\{\kthreshold_{i,k_i^{max}}\}$ for adaptation of $k_i$ is found in a similar fashion.

\begin{algorithm}[t]
\KwIn{$\overline{q} = 0$. When {\tt request} arrives}
	$q \leftarrow $ queue length upon arrival of {\tt request}\;
	$i\leftarrow$ class that {\tt request} belongs to\;
	$\overline{q} \leftarrow \alpha \overline{q} + (1-\alpha)q$\;
	Find $k\le k_i^{max}$ such that $\overline{q}\in [\nthreshold_{i,k+1},\nthreshold_{i,k})$\;
	Find $n\le n_i^{max}$ such that $\overline{q}\in [\nthreshold_{i,n+1},\nthreshold_{i,n})$\;
	$n \leftarrow \min(r_i^{max}k, n)$\;	\label{step:double-check-n}
	Serve {\tt request} with an $(n,k)$ code when it becomes HoL\;
	\caption{\ourproposal (Throughput Optimal FEC Cloud)}
	\label{alg:TOFEC}
\end{algorithm}

The adaptation mechanism of \ourproposal is summarized in pseudo-codes as Algorithm \ref{alg:TOFEC}.
Note that in Step \ref{step:double-check-n} we reduce $n$ to $r_i^{max}k$ if the redundancy ratio of the code chosen in the previous steps is higher than $r_i^{max}$ -- the maximum allowed redundancy ratio for class $i$. 
Also, instead of comparing $q$ directly with the thresholds, we compare an exponential moving average 
$\overline{q} = \alpha \overline{q} + (1-\alpha)q$,
with a memory factor $0\le \alpha \le 1$, against the thresholds to determine $n$ and $k$. The moving average is used to mitigate the transient variation in queue length so that $n$ and $k$ will not change too frequently. It is obvious that we only need to set $\alpha=0$ in order to use instantaneous queue length $q$ for the adaptation since in this case $\overline{q}=q$.

It is worth pointing out that \ourproposal's threshold based adaptation is 
\begin{enumerate}
	\item Independent of $\compVec$: The thresholds for each class is computed a priori without any knowledge or assumption of $\compVec$. Once computed, the thresholds can be reused for all realizations of different $\compVec$, even if $\compVec$ is time-varying;
	\item Independent across classes: For a class $i$, computation of its thresholds  require knowledge of neither the number nor the delay parameters of other classes. The adaptation of class $i$ is also independent of those of the other classes.
\end{enumerate}
These two properties of independence are direct result of the implication of Theorem \ref{thm:optimal} we discussed before.
Thanks to these nice properties, it is very easy in \ourproposal to add support for a new class in an incremental fashion: simply compute the thresholds for the new class, leaving the thresholds for the existing, say $m$, classes untouched. The old and new thresholds together will then produce the optimal choice of codes for the incremented set of $m+1$ classes.

\section{Evaluation}
\label{sec:evaluation}

We now demonstrate the benefits of \ourproposal 's adaptation mechanism. 
We evaluate \ourproposal's adaptation strategy and show that is outperforms static strategies with both constant and changing workloads, as well as a simple greedy heuristic that will be introduced later.

\begin{figure*}[!t]
\centering
\null\hfill
	\subfigure[Mean Delay]{
		\label{fig:read:ave}
		\includegraphics[width=\onewidth]{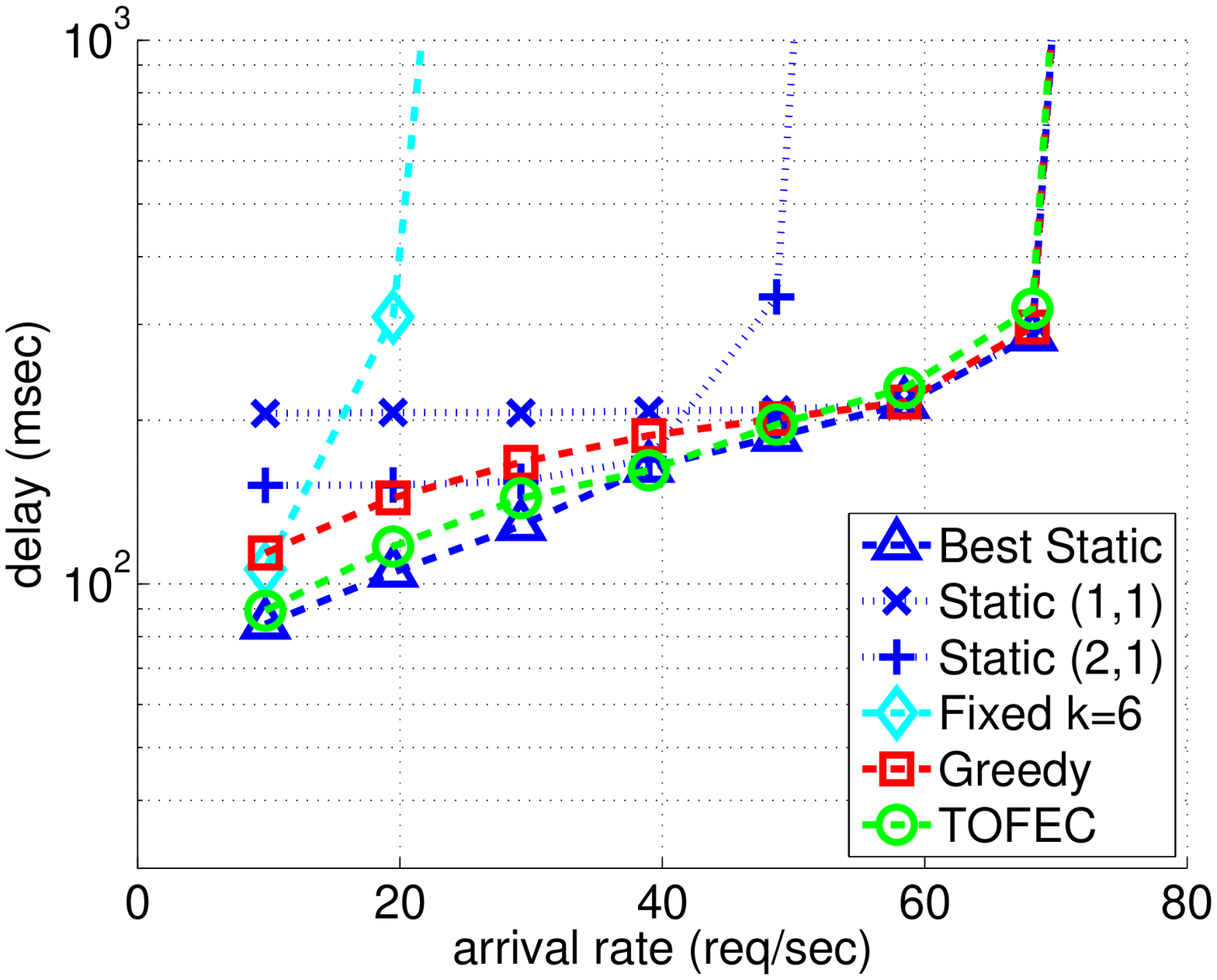}
	}%
\hfill
	\subfigure[Median Delay]{
		\label{fig:read:med}
		\includegraphics[width=\onewidth]{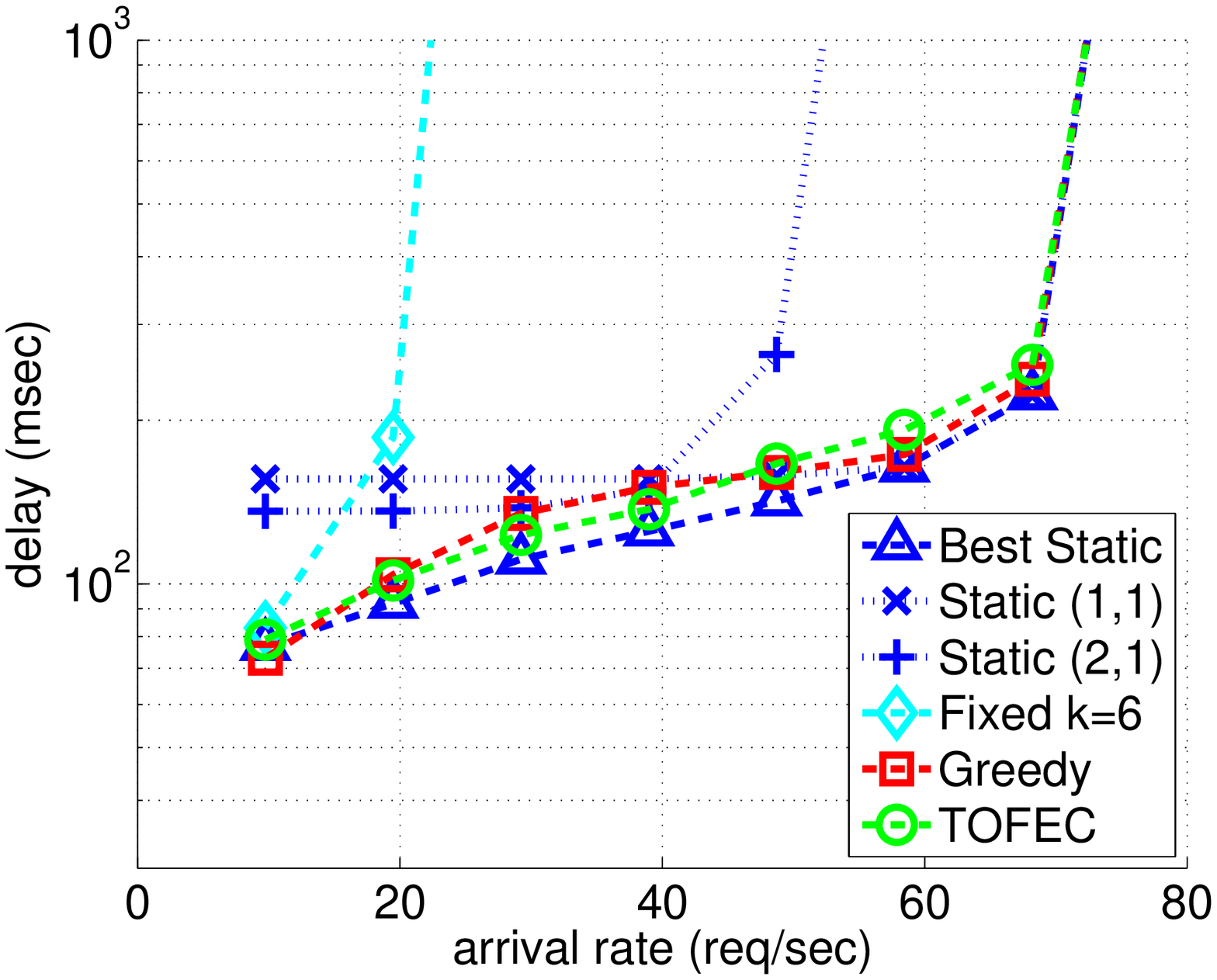}
	}
\hfill\null
\\
\null\hfill
	\subfigure[90th Percentile Delay]{
		\label{fig:read:9}
		\includegraphics[width=\onewidth]{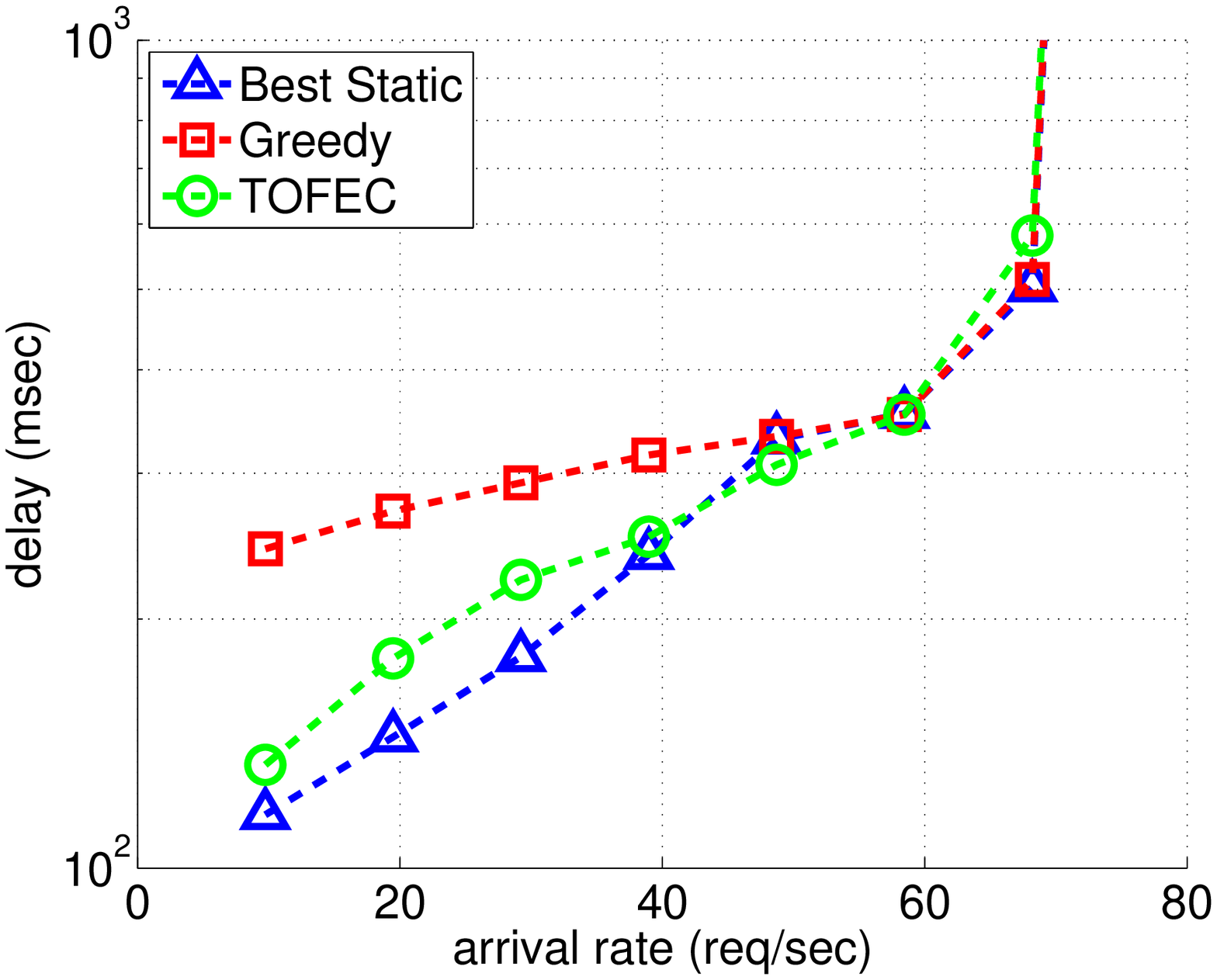}
	}
\hfill
	\subfigure[99th Percentile Delay]{
		\label{fig:read:99}
		\includegraphics[width=\onewidth]{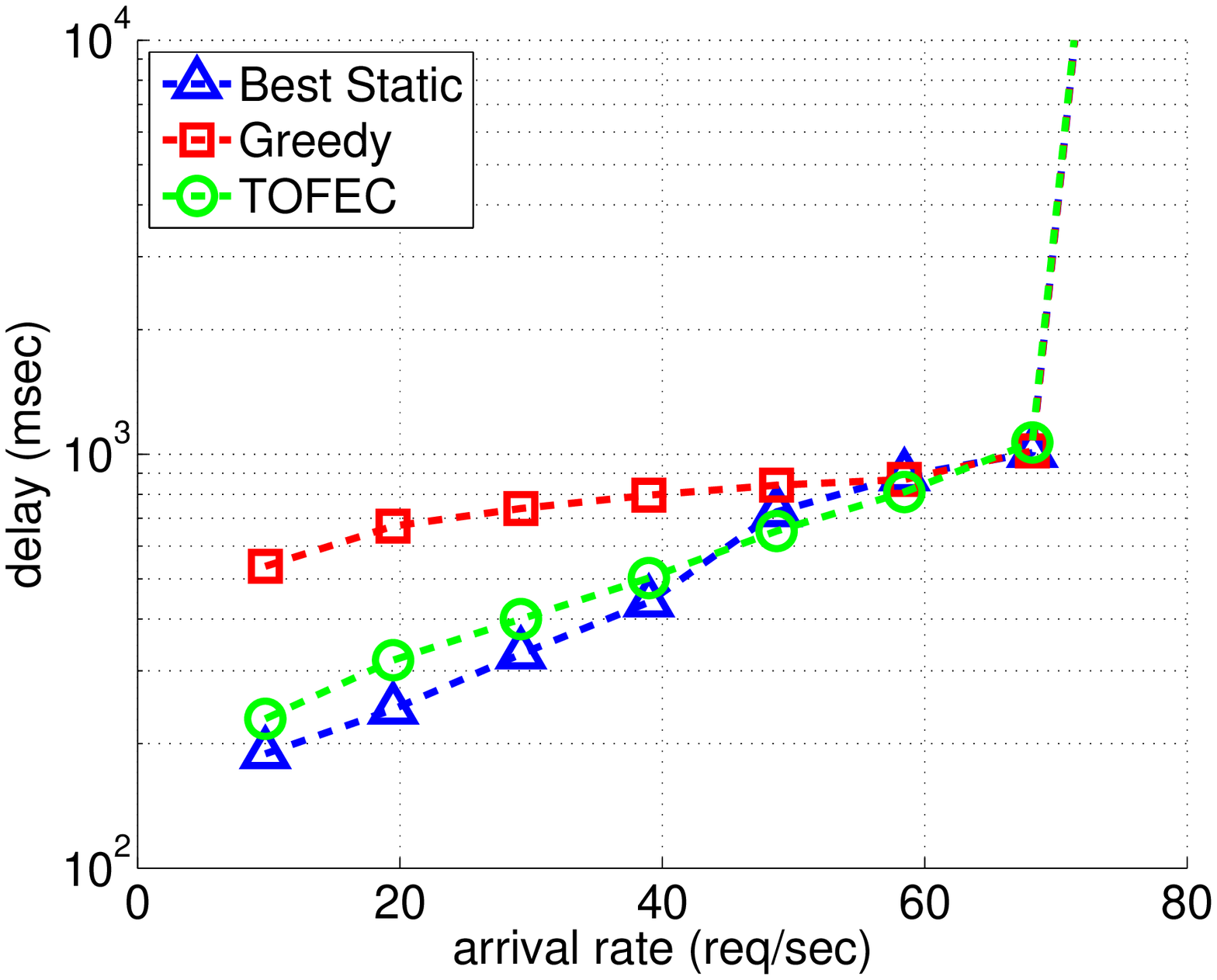}
	}
\hfill\null
\caption{Delay performance in read only scenario}
\label{fig:read}
\end{figure*}
\subsection{Simulation Setup}
\label{ssec:eva:setup}

We conducted trace-driven simulations for performance evaluation for both single-class and multi-class scenarios with both read and write requests of different file sizes. 
Due to lack of space, we only show results for the scenario with one class {\tt (read,3MB)}. But we must emphasize that it is representative enough so that the findings to be discussed in this section are valid for other settings (different file sizes, write requests, and multiple classes). 
We assume that the system supports up to $L=16$ simultaneously active threads. 
We set the maximum code dimension and redundancy ratio to be $k^{max} = 6$ and $r^{max} = 2$, because we observe negligible gain in service delay beyond this chunking and redundancy level from our measurements. 
We use traces collected in May and June 2013 in availability region ``North California''. In order to compute the thresholds for \ourproposal, we need estimations of the delay parameters $\{\fixedDelta,\linearDelta,\fixedExp,\linearExp\}$. For this, we first filter out the worst 10\% task delays in the traces, then we compute the delay parameters from the least squares linear approximation for the mean and standard deviation of the remaining task delays. We use memory factor $\alpha = 0.99$ in \ourproposal.

In addition to the static strategies, we develop a simple {\em Greedy} heuristic strategy for the purpose of comparison. Unlike the adaptive strategy in \ourproposal, Greedy does not require prior-knowledge of the distribution of task delays, yet as the results will reveal, it achieves a competitive mean delay performance. In Greedy, the code to be used to serve a request in class $i$ is determined by the number of idle threads upon its arrival: suppose there are $l$ idle threads, then
$$
k_i = 
\begin{cases}
1, & \mbox{if } l=0,\\
\min(k_i^{max},l), & \mbox{otherwise};
\end{cases}
$$
and similarly
$$
n_i = 
\begin{cases}
1, & \mbox{if } l=0,\\
\min(r_i^{max}k_i,l), & \mbox{otherwise}.
\end{cases}
$$
The idea of Greedy is to first maximize the level of chunking with the idle threads available, then increase the redundancy ratio as long as there are idle threads remaining.


\subsection{Throughput-Delay Trade-Off}
Fig.\ref{fig:read} shows the mean, median, 90th percentile and 99th percentile delays of  \ourproposal and Greedy with Poisson arrivals at different arrival rates of $\lambda$. We also run simulations with static strategies for all possible combinations of $(n,k)$ at every arrival rate. In a brute-force fashion, we find the best mean, median, 90th and 99th percentile delays achieved with static strategies and use them as the baseline. 
Fig.\ref{fig:read:ave} and Fig.\ref{fig:read:med} also plot the mean and median delay performance of the basic static strategy with no chunking and no replication, i.e., $(1,1)$ code; the simple replication static strategy with a $(2,1)$ code; and the backlog-based adaptive strategy from \cite{fastcloud} with fixed code dimension $k=6$ and $n\le 12$.

As we can see, 
both \ourproposal and Greedy successfully support the full capacity region -- the one supported by basic static -- while achieving almost optimal mean and median delays throughout the full capacity region.
At light workload, \ourproposal delivers about $2.5\times$ improvement in mean delay when compared with the basic static strategy, and about $2\times$ when compared with simple replication (from 205ms and 151ms to 84ms). It also reduces the median delay by about $2\times$ from that of basic and simple replication (from 156ms and 138ms to 74ms). Meanwhile Greedy achieves about $2\times$ improvement in both mean (89ms) and median delays (79ms) over basic. 

With heavier workload, both \ourproposal and Greedy successfully adapt their codes to keep track with the best static strategies, in terms of mean and median delays. It is clear from the figures that both \ourproposal and Greedy achieve our primary goal of retaining full system capacity, as supported by basic static strategy. On the contrary, although simple replication has slightly better mean and median delays than basic under light workload, it fails to support arrival rates beyond 70\% of the capacity of basic.
Meanwhile, the adaptive strategy from \cite{fastcloud} with fixed code dimension $k=6$ can only support less than 30\% of the original capacity region, although it achieves the best delay at very light workload.

\comment{
\begin{figure}[t]
	\subfigure[Average $k$]{
		\label{fig:read:codeDim}
		\includegraphics[width = \twowidth]{read_codeDim}
	}%
	\subfigure[Average $n$]{
		\label{fig:read:codeLength}
		\includegraphics[width = \twowidth]{read_codeLength}
	}
\vspace{\shrinkbeforecaption}
\caption{Delay performance in read only scenario}
\label{fig:read:code}
\vspace{\shrinkaftercaption}
\end{figure}
}

While the two adaptive strategies have similar performance in mean and median, \ourproposal outperforms Greedy significantly at high percentiles. As Fig.\ref{fig:read:9} and Fig.\ref{fig:read:99} demonstrate, \ourproposal is on a par with the best static strategies at the 90th and 99th percentile delays throughout the whole capacity region. On the other hand, Greedy fails to keep track of the best static performance at lower arrival rates. At light workload, \ourproposal 's is over $2\times$ and $2.5\times$ better than Greedy at the 90th and 99th percentiles. Less interesting is the case with heavy workload when the system is capacity-limited. Hence both strategies converge to the basic static strategy using mostly $(1,1)$ code, which is optimal at this regime.

\subsection{Delay Variation and Choice of Codes}

\comment{
In Fig.\ref{fig:read:codeDim} and Fig.\ref{fig:read:codeLength} we plot the average code dimension $k$ and code length $n$ in \ourproposal and Greedy, as well as the code used by the static strategies that produce the best mean delay at different arrival rates. These figures again confirm that the adaptation algorithms of \ourproposal and Greedy are working as they are designed to: the average code dimension and length both match with the best static strategies quite well. Greedy is a bit too aggressive in choosing code dimension when compared with the best static strategy: the average code dimension of Greedy is always at least as large as that of the best static strategy. On the contrary,  \ourproposal's choice of code dimension turns out to be a better interpolation of the best static strategy.
}

\begin{figure}[t]
\centering
\includegraphics[width=\onewidth]{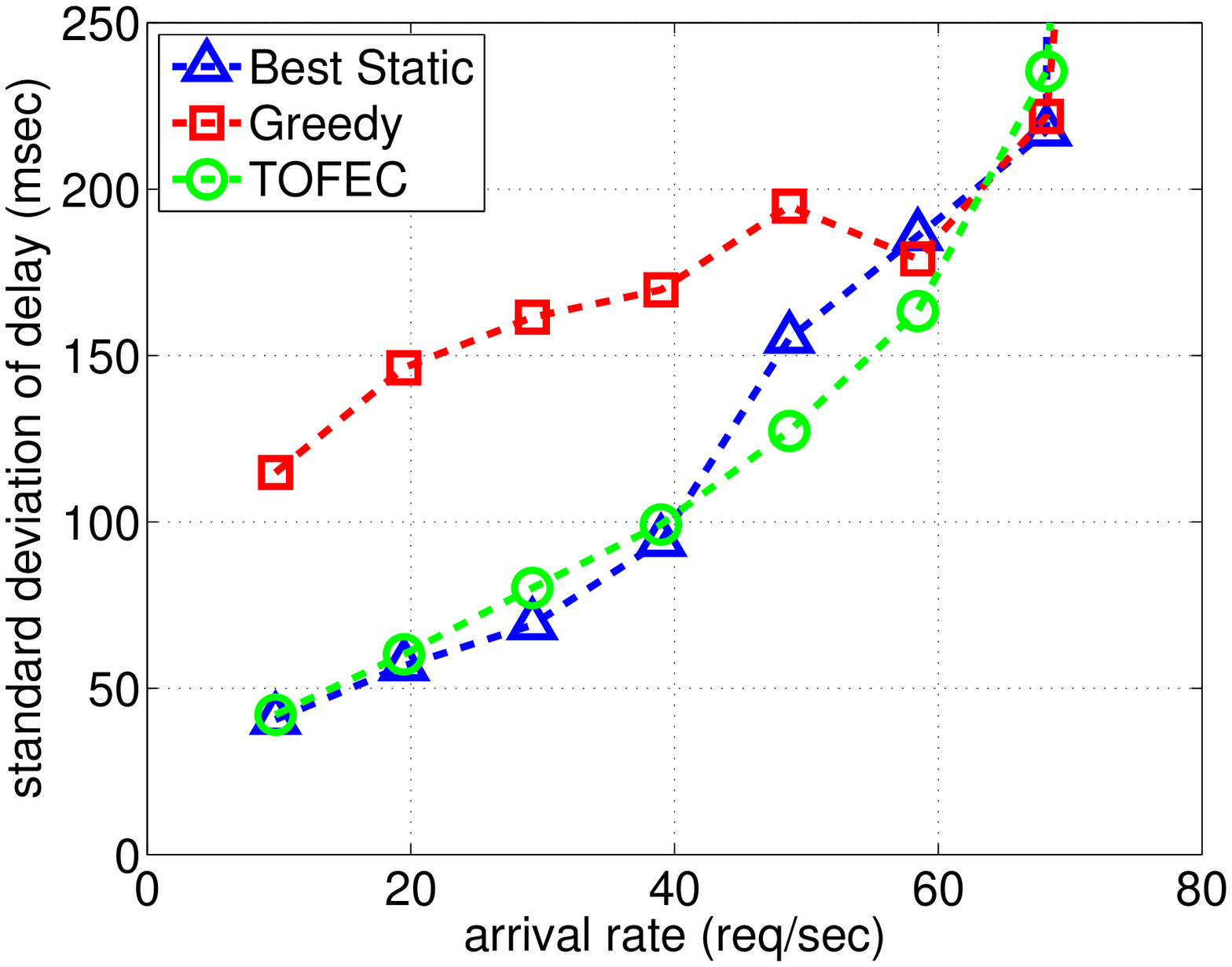}
\caption{Comparison of standard deviation}
\label{fig:read:std}
\end{figure}

We further compare the standard deviation (STD) of \ourproposal, Greedy and the best static strategy. STD is a very important performance metric because it directly relates to whether customers can receive consistent QoS. In certain applications, such as video streaming, maintaining low STD in delay can be even more critical than achieving low mean delay. As we can see in Fig.\ref{fig:read:std}, for the region of interest with light to medium workload, \ourproposal delivers $2\times$ to $3\times$ lower STD than Greedy does. Moreover, in spite of its dynamic adaptive nature, \ourproposal in fact matches with the best static strategy very well throughout the full capacity region. This suggests the code choice in \ourproposal indeed converges to the optimal.

\begin{figure}[t]
\centering
\includegraphics[width=\onewidth]{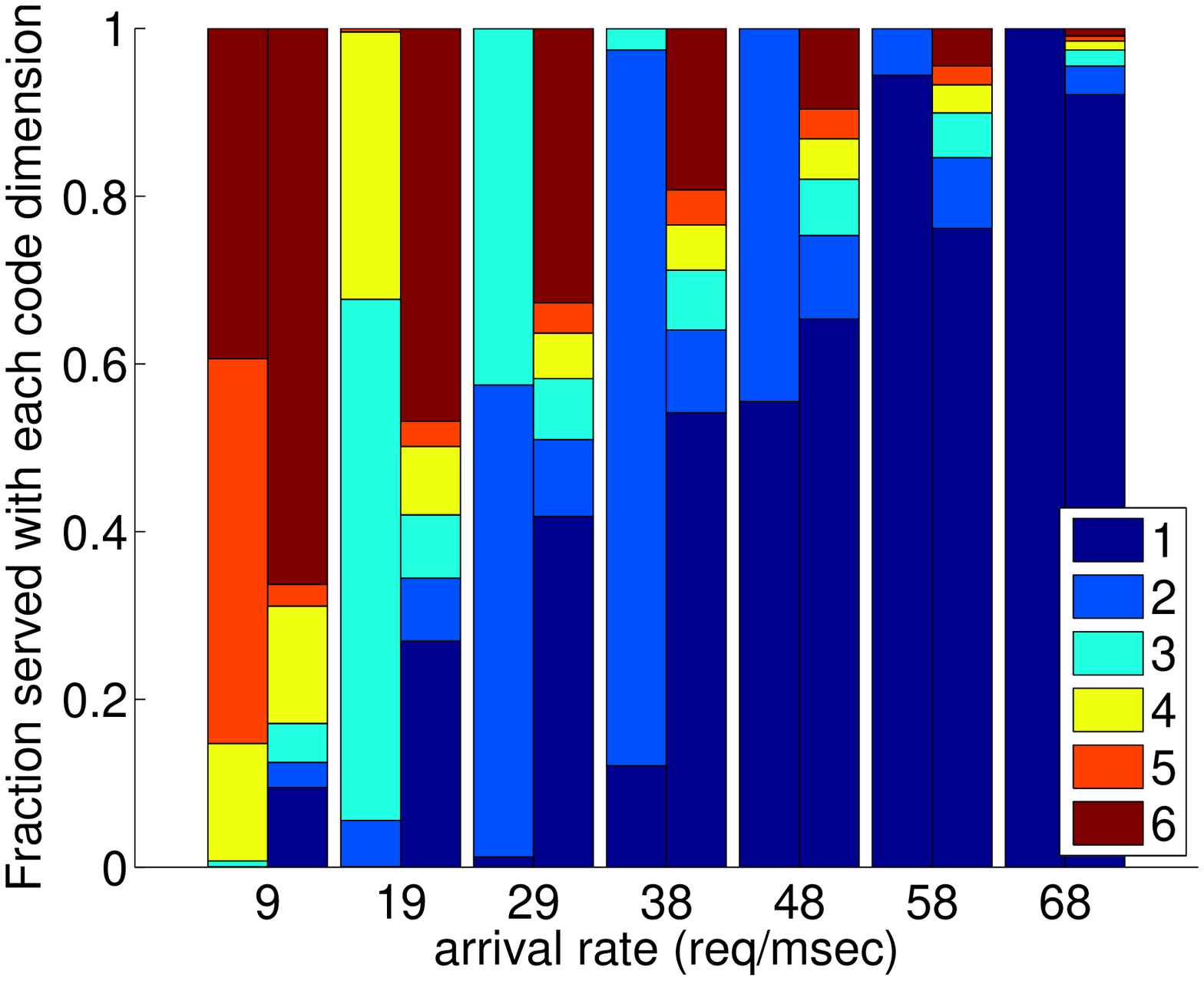}
\caption{Composition of $k$. Left: \ourproposal, Right: Greedy}
\label{fig:read:composition}
\end{figure}

The convergence to optimal becomes more obvious when we look into the fraction of requests served by each choice of code. In Fig.\ref{fig:read:composition} we plot the compositions of requests served by different code dimension $k$'s. 
At each arrival rate, the two bars represent \ourproposal and Greedy. For each bar, blocks in different colors represent the fraction of requests served with code dimension
1 through 6, from bottom to top. \ourproposal's choice of $k$ demonstrates a high concentration around the optimal value: at all arrival rate, over 80\%  requests are served by 2 neighboring values of $k$ around the optimal, and this fraction quickly diminishes to 0 for codes further from the optimal. Moreover, as arrival rate varies from low to high, \ourproposal's choice of $k$ transitions quite smoothly as $(5,6) \rightarrow (3,4) \rightarrow  (2,3) \rightarrow (1,2)$ and eventually converges to a single value $1$ as workload approaches system capacity.

On the contrary, Greedy tends to round-robin across all possible choices of $k$ and majority of requests are served by either $k=1$ or $6$. So Greedy is effectively alternating between the two extremes of no chunking and very high chunking,  instead of staying around the optimal. 
Such ``all or nothing'' behavior results in the $2\times$ to $3\times$ worse STD shown in Fig.\ref{fig:read:std}. So \ourproposal provides much better QoS guarantee.



\begin{figure*}[!t]
\centering
	\subfigure[Sampled delay]{
		\label{fig:read:changeArrival}
		\includegraphics[width=\threewidth]{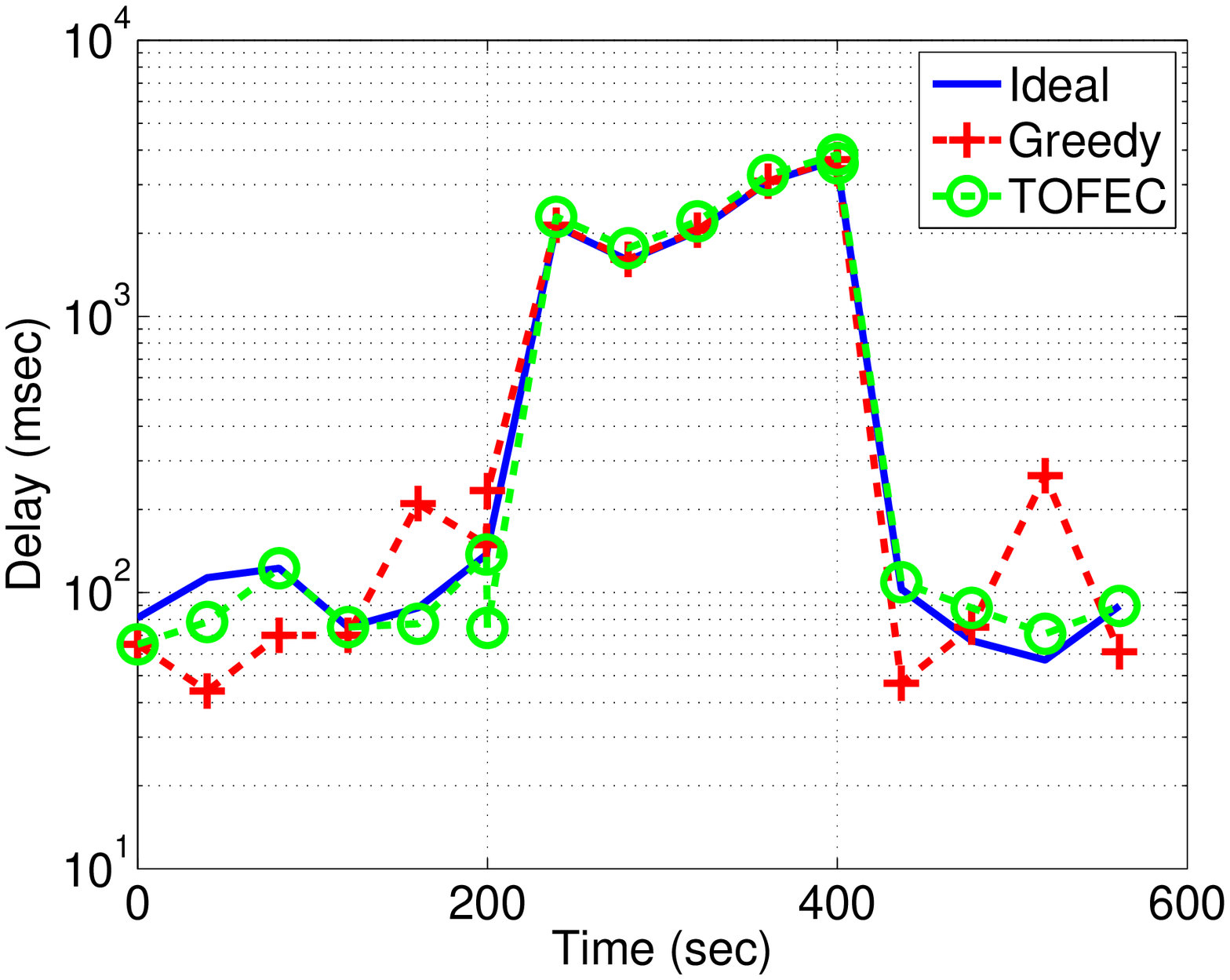}
	}
\hfill
	\subfigure[Sampled code dimension $k$]{
		\label{fig:read:changeArrival:codeDim}
		\includegraphics[width=\threewidth]{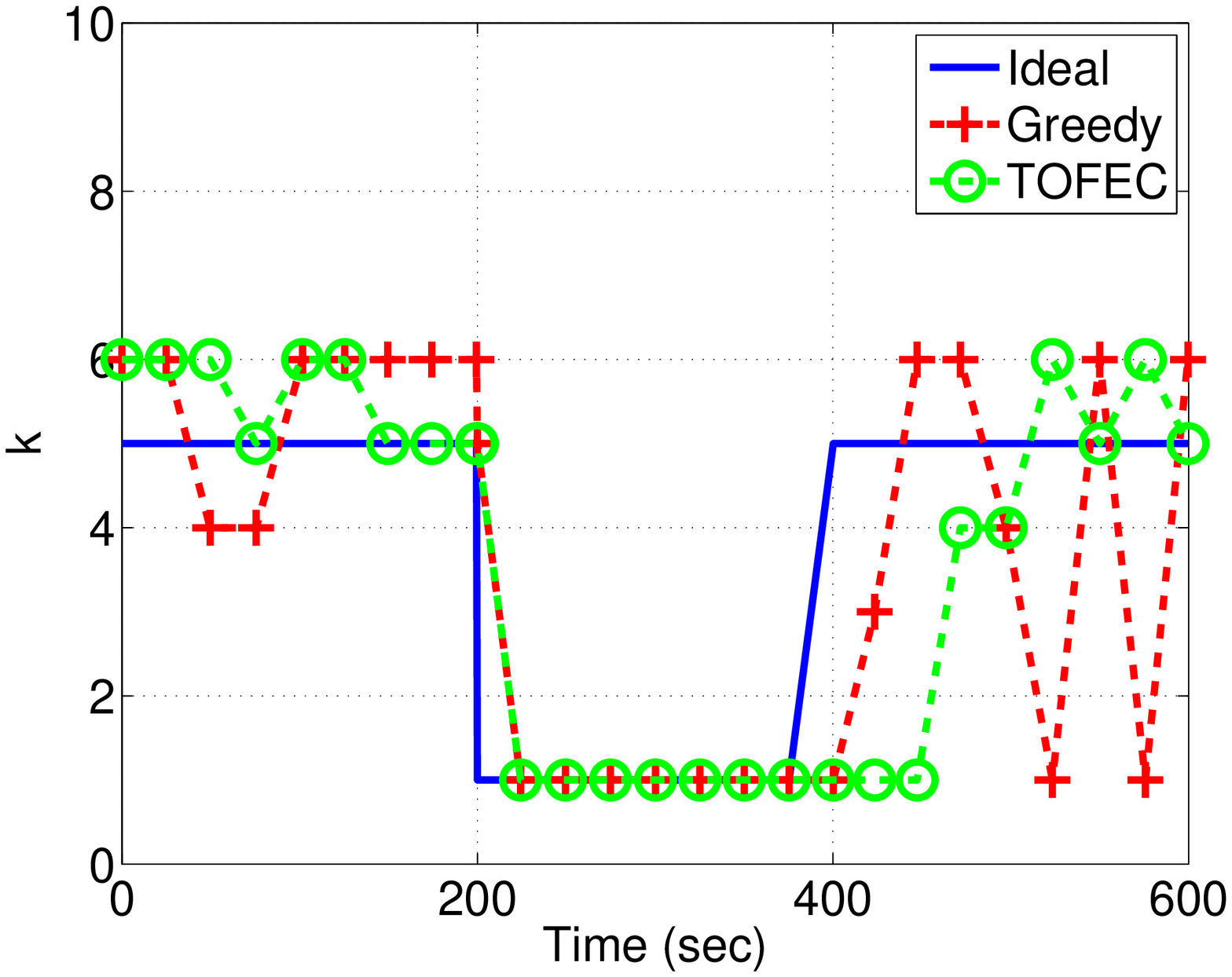}
	}
\hfill
	\subfigure[Zoom-in delays]{
		\label{fig:read:changeArrival:zoomin}
		\includegraphics[width=\threewidth]{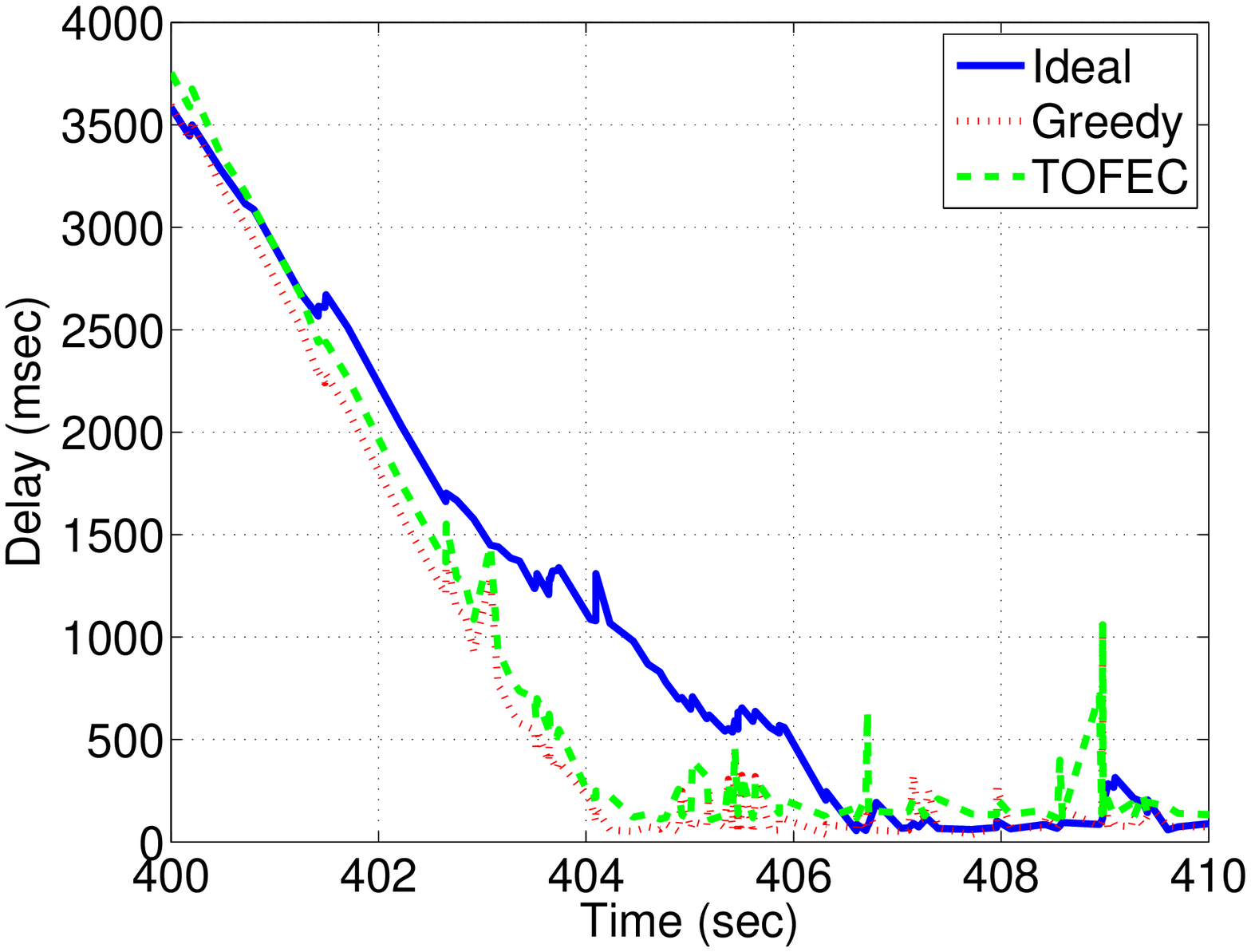}
	}
\caption{Adaptation to changing workload}
\label{fig:adapt}
\end{figure*}

\comment{
\begin{figure}[!t]
\centering
\includegraphics[width = \onewidth]{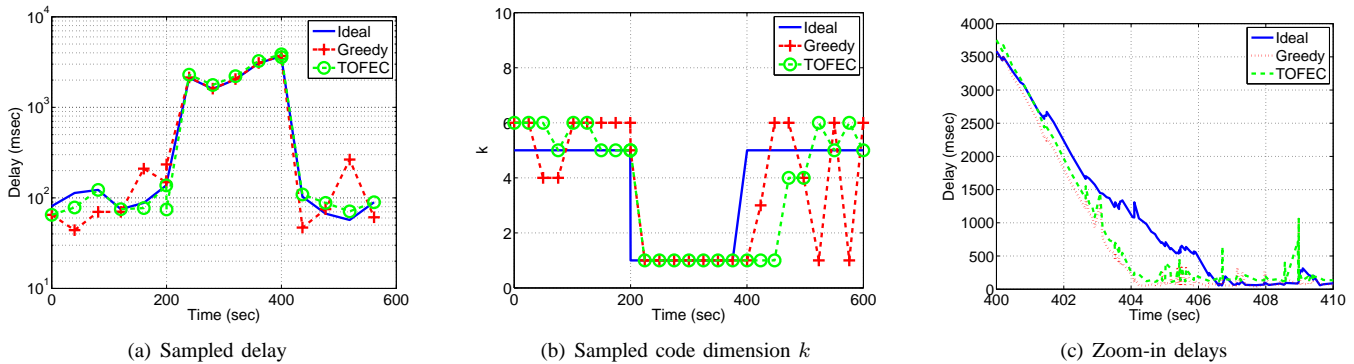}
\caption{Adaptation to changing workload}
\label{fig:read:changeArrival}
\end{figure}
}

\subsection{Adapting to Changing Workload}
We further examine how well the two adaptive strategies adjust to changes in workload. 
In Fig.\ref{fig:adapt} we plot the total delay experienced by requests arriving at different times within a 600-second period, as well as the choice of code in the same period. The 600 seconds is divided into 3 phases, each lasts 200 seconds.
The arrival rate is 10 request/second in phases 1 and 3, and 80 request/second (slightly $>\CapFull$) in phase 2. The corresponding optimal choices of codes $(n,k)$ are $(10,5)$ for phases 1 and 3, and $(1,1)$ for phase 2. 
For the purpose of comparison, we also implement an ``Ideal'' rate-driven strategy that has perfect knowledge of the arrival rate of each phase and picks the optimal code accordingly as the baseline.
We can see that both \ourproposal and Greedy are quite agile to changes in arrival rate and quickly converge to a good composition of codes that delivers optimal mean delays within each phase,  comparable to that of Ideal.

From Fig.\ref{fig:read:changeArrival:codeDim} we can further observe that \ourproposal is especially responsive in face of workload surge (from phase 1 to 2). This is because the suddenly increased arrival rate immediately builds up a large backlog, which in turn forces \ourproposal to pick a code with the smallest $k=1$. When the arrival drops (from phase 2 to 3), instead of immediately switching back to codes with $k=5$, \ourproposal gradually transitions to the optimal value of $k=5$. Such ``smoothening'' behavior when workload reduces is actually beneficial. This is because the request queue has been built up during the preceding period of heavy workload. So, if $k$ is set to 5 right after arrival rate drops, it will produce a throughput so low that it takes a much longer time to reduce the queue length to the desired level, and requests arrive during this period will suffer from long queueing delay even though they are being served with the optimal code. On the other hand, \ourproposal's queue length driven adaptation sticks with smaller $k$, which delivers higher throughput, to drain the queue much faster to the desired level.  As we can see in Fig.\ref{fig:read:changeArrival:zoomin}, which plots the delay traces for requests arrive in the first 10 seconds of phase 3, \ourproposal and Greedy both reduce their delay to optimal almost $1.8\times$  faster than Ideal does after workload decreases. This is another advantage of using queue length instead of arrival rate to drive code adaptation.

We can also see that \ourproposal's choice of code is much more stable than that of Greedy. 
While \ourproposal shows little variation around the optimal in each phase, Greedy keeps oscillating between $k=1$ and $k=6$ when the optimal is 1! This is consistent with the ``all or nothing'' behavior of Greedy observed in Fig.\ref{fig:read:composition}.

\section{Related Work}
\label{sec:related}

FEC in connection with multiple paths and/or multiple servers is a well investigated topic in the literature \cite{VickySharmaMPLOT,EminGabrielyanFEC,JohnByersAccessing,RSaadEvaluating}. However, there is very little attention devoted to the queueing delays. FEC in the context of network coding or coded scheduling has also been a popular topic from the perspectives of throughput (or network utility) maximization and throughput vs. service delay trade-offs \cite{Eryilmaz:2008:DTG:2263482.2273567,Yeownetworkcoding,Theodorosnetworkcoding, KozatScheduling}. Although some incorporate queuing delay analysis, the treatment is largely for broadcast wireless channels with quite different system characteristics and constraints.
FEC has also been extensively studied in the context of distributed storage from the points of high durability and availability while attaining high storage efficiency \cite{Dimakis:2010:NCD:1861840.1861868,Rodrigues_highavailability,Li:2010:TDR:1833515.1833884}. 

Authors of \cite{Longbocodeingincloud} conducted theoretical study of cloud storage systems using FEC in a similar fashion as we did in our work \cite{fastcloud}. Given that exact mathematical analysis of the general case is very difficult, authors of \cite{Longbocodeingincloud} considered a very simple case with a fixed code of $k=2$ tasks. Shah et al. \cite{MDS-queue} generalize the results from \cite{Longbocodeingincloud} to $k>2$. Both works rely on the assumption of exponential task delays, which hardly captures the reality. Therefore, some of their theoretical results cannot be applied in practice.
For example, under the assumption of exponential task delays, Shah et al. have proved that it is optimal to always use the largest $n$ possible throughout the full capacity region $\CapFull$,
contradicting with simulation results using real-world measurements in \cite{fastcloud} and this paper. 

\section{Conclusion}
\label{sec:conclusion}
This paper presents the first set of solutions for achieving the optimal throughput-delay trade-off for scalable key-value storage access using erasure codes with variable chunk sizing and rate adaptation. We establish the viability of this approach through extensive measurement study over the popular public cloud storage service Amazon S3. We develop two adaptation strategies: \ourproposal and Greedy.  \ourproposal monitors the local backlog and compares it against a set of thresholds to dynamically determine the optimal code length and dimension. Our trace-driven simulation shows that \ourproposal is on a par with the best static strategy in terms of mean, median, 90th, and 99th percentile delays, as well as delay variation. To compute the thresholds, \ourproposal requires knowledge of the mean and variance of cloud storage access delays, which is usually obtained by maintaining a log of delay traces. On the other hand, Greedy does not require any knowledge of the delay profile or logging but is able to achieve mean and median delays comparable to those of \ourproposal. However, it falls short in important QoS metrics such as higher percentile delays and variation. It is part of our ongoing work to develop a strategy that matches \ourproposal's high percentile delay performance without prior knowledge and logging.

\bibliographystyle{IEEEtran}
\bibliography{PaperList}

\appendix[Proof of Theorem \ref{thm:optimal}]
\label{sec:appendix}

\begin{IEEEproof}
\comment{
It is easy to verify that within the feasible region
\begin{align}
\frac{\partial D_q}{\partial \aveUsage}
&=\frac{L}{(L-\lambda \aveUsage)^2} - \frac{1}{L},
\\
\frac{\partial \aveUsage}{\partial k_i}
& = p_i (\fixedDelta_i r_i + \fixedExp_i ),
\\
\frac{\partial \aveUsage}{\partial r_i} 
&= p_i (\fixedDelta_i k_i + \linearDelta_i J_i ),
\\
\frac{\partial D_{s,i}}{\partial k_i}
&= 
-\frac{J_i}{k_i^2}\left(\linearDelta_i + \linearExp_i\ln\frac{r_i}{r_i-1}\right),
\\
\frac{\partial D_{s,i}}{\partial r_i}
&= 
-\left(\fixedExp_i + \frac{\linearExp_i J_i}{k_i}\right)\frac{1}{r_i(r_i-1)}.
\end{align}
}
It is easy to verify that the objective function \eqref{eq:optimization} is continuous and differentiable everywhere within the feasible region and the partial derivatives are 
\begin{align}
\frac{\partial (\ast)}{\partial k_i} 
=& \left(\frac{L}{(L-\lambda \aveUsage)^2} - \frac{1}{L}\right) p_i (\fixedDelta_i r_i + \fixedExp_i ) 
\nonumber\\ 
& - p_i \frac{J_i}{k_i^2}\left(\linearDelta_i + \linearExp_i\ln\frac{r_i}{r_i-1}\right)
\label{eq:partial:k}
\end{align}
and 
\begin{align}
\frac{\partial (\ast)}{\partial r_i} 
=& \left(\frac{L}{(L-\lambda \aveUsage)^2} - \frac{1}{L}\right) p_i (\fixedDelta_i k_i + \linearDelta_i J_i )
\nonumber\\
& - p_i\left(\fixedExp_i + \frac{\linearExp_i J_i}{k_i}\right)\frac{1}{r_i(r_i-1)}
\label{eq:partial:r}
\end{align}

Notice that for the whole the feasible region including the boundary, \eqref{eq:optimization} is always lower bounded by 0. So there must exist at least one global optimal solution \optSolution that minimizes \eqref{eq:optimization}. Moreover, \eqref{eq:optimization} goes to $\infty$ if and only if the operating point $(\dimVec,\rateVec)$ approaches the boundary, i.e., $k_i \rightarrow 0$, or $r_i \rightarrow 1$, or $\lambda \rightarrow \CapSta(\compVec,\dimVec,\rateVec)$. Since \eqref{eq:optimization} is $\infty$ for the whole boundary, the global optimal \optSolution must reside strictly within the feasible region. As a result, both Eq.\ref{eq:partial:k} and Eq.\ref{eq:partial:r} must evaluate to 0 at \optSolution. 
In the subsequent discussion, we prove that Eq.\ref{eq:partial:k}~=~0 and Eq.\ref{eq:partial:r}~=~0 has an unique solution within the feasible region. As a result, \optSolution equals to this solution and is also unique.

From Eq.\ref{eq:partial:r}~=~0 we have:
\begin{equation}
\left(\frac{L}{(L-\lambda \aveUsage)^2} - \frac{1}{L}\right)
= \frac{ \fixedExp_i k_i + \linearExp_i J_i}{k_i r_i (r_i-1) (\fixedDelta_i k_i + \linearDelta_i J_i)}.
\label{eq:opt:normArrival:app}
\end{equation}
Plugging the above into Eq.\ref{eq:partial:k}, we have:
\begin{equation}
\frac{k_i(\fixedExp_i k_i  + \linearExp J_i)}{\fixedDelta_i k_i + \linearDelta_i J_i} 
= \frac{J_i r_i(r_i-1)}{\fixedDelta_i r_i + \fixedExp_i}\left(\linearDelta_i + \linearExp_i\ln\frac{r_i}{r_i-1}\right).
\label{eq:opt:code:app}
\end{equation}

Notice that if $r_i$ is fixed, Eq.\ref{eq:opt:code:app} is in fact a quadratic equation of $k_i$. 
Let $\FR_i(r_i)$ (or $\FR_i$ for short) be the right hand side of Eq.\ref{eq:opt:code:app}. Then we have
\begin{align}
k_i &= \frac{\fixedDelta_i \FR_i - \linearExp_i J_i + \sqrt{(\fixedDelta_i \FR_i - \linearExp_i J_i)^2 + 4\fixedExp_i \linearDelta_i J_i \FR_i }}{2\fixedExp_i}
\\
&\triangleq \KR_i(r_i).
\end{align}
We do not consider the other solution to Eq.\ref{eq:opt:code:app} because it is always $\le 0$.
It is easy to verify that $\KR_i$ is a strictly increasing function of $r_i$ within the feasible region. 

Substituting $k_i$ with $\KR_i(r_i)$, the right hand side of Eq.\ref{eq:opt:normArrival:app} can be written as a function $\pi_i(r_i)$, which can be shown to be strictly decreasing. Notice that Eq.\ref{eq:opt:normArrival:app} must be satisfied for all $i$ and the left hand side remains unchanged. Then
\begin{equation}
\pi_i(r_i) = \pi_j(r_j),~\forall i,j.
\label{eq:pi:constant}
\end{equation}
Note that $\pi_i$ and $\pi_j$ are strictly decreasing functions of $r_i$ and $r_j$, respectively. This means that there is a one-to-one mapping between any $r_i$ and $r_j$ at the optimal solutions, and $r_j$ is a strictly increasing function of $r_i$, namely $r_j = \RR_{j,i}(r_i)$. 

Now with $r_j = \RR_{j,i}(r_i)$ and $k_j = \KR_j(r_j) = \KR_j(\RR_{j,i}(r_i))$, Eq.\ref{eq:opt:normArrival:app} becomes a equation that contains only one variable $r_i$. It is then not hard to show that for any given $\lambda$ and $\compVec$, the left hand side of Eq.\ref{eq:opt:normArrival:app} is a strictly increasing function of $r_i$, while the right hand side is $\pi_i(r_i)$, which is strictly decreasing. As a result, these two functions can equal for at most one value of $r_i$. In other words, equations Eq.\ref{eq:opt:code:app} and Eq.\ref{eq:opt:normArrival:app} have at most one solution. The existence of a solution to these equations is guaranteed by the existence of \optSolution, so it must be unique. This completes the proof.
\end{IEEEproof}

\begin{IEEEbiography}[{\includegraphics[width=1in,height=1.25in,clip,keepaspectratio]{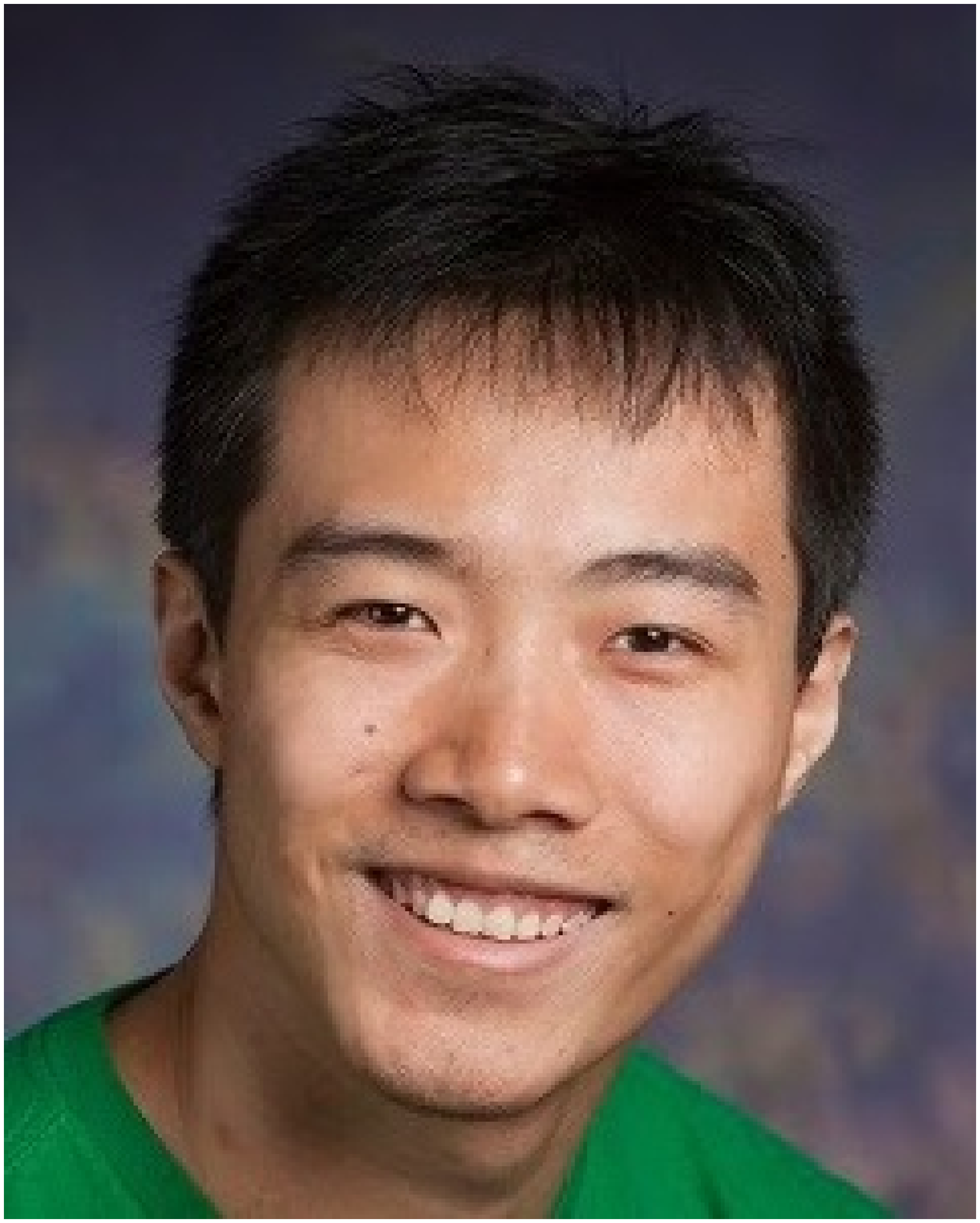}}]{Guanfeng Liang}
(S'06-M'12)
received his B.E. degree from University of Science and Technology of China, Hefei, Anhui, China, in 2004, M.A.Sc. degree in Electrical and Computer Engineering from  University of Toronto, Canada, in 2007, and Ph.D. degree in Electrical and Computer Engineering from the University of Illinois at Urbana-Chanpaign, in 2012. 
He currently works with DOCOMO Innovations (formerly DOCOMO USA Labs), Palo Alto, CA, as a Research Engineer. 
\end{IEEEbiography}

\begin{IEEEbiography}[{\includegraphics[width=1in,height=1.25in,clip,keepaspectratio]{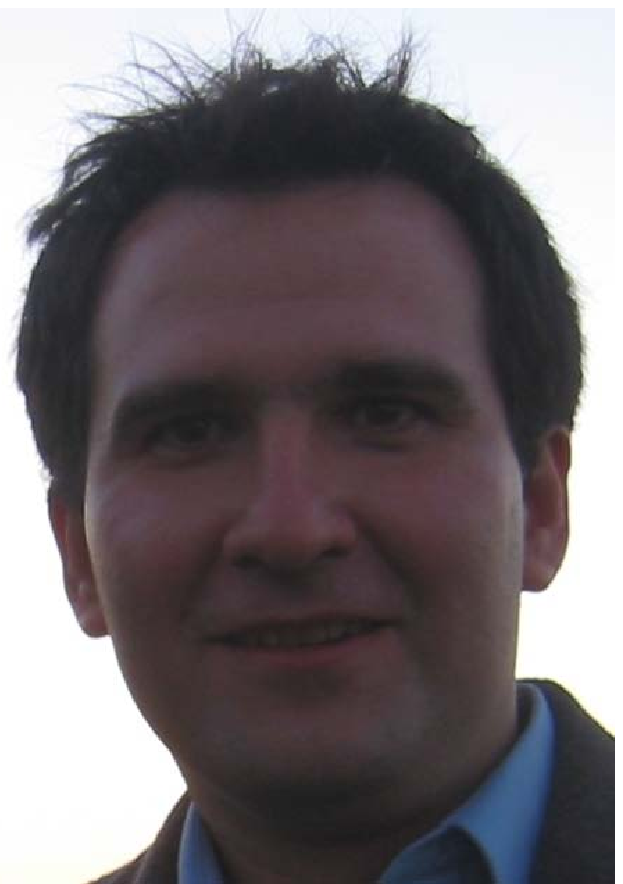}}]{Ula\c{s}~C.~Kozat} 
(S’97-M’04-SM’10) received his
B.Sc. degree in Electrical and Electronics Engineering
from Bilkent University, Ankara, Turkey, in 1997,
M.Sc. degree in Electrical Engineering from the
George Washington University, Washington, DC, in
1999, and Ph.D. degree in Electrical and Computer
Engineering from the University of Maryland,
College Park, in 2004. He currently works at DOCOMO Innovations (formerly DOCOMO USA Labs), Palo Alto, CA, as
a Principal Researcher.
\end{IEEEbiography}
\end{document}